\documentclass[twocolumn,twocolappendix]{aastex63}

\usepackage{xspace}
\usepackage{natbib}
\usepackage{url}

\newcommand\wfirst{{\textit{WFIRST}}\xspace}
\newcommand\romanst{{\textit{Roman}}\xspace}
\newcommand\gulls{{\sc GULLS}\xspace}
\newcommand{\tsub}[2]{{#1}_\textrm{\tiny{#2}}}
\newcommand\tE{{$t_{\textrm{\tiny E}}$\xspace}}
\defcitealias{penny2019}{Paper I}
\newcommand{\epsScaleFactor}{1.17}

\graphicspath{{./}{figures/}}

\received{\today}
\revised{}
\accepted{}

\submitjournal{ApJ}

\shorttitle{\romanst~FFPs}
\shortauthors{Johnson et al.}

\begin{document}

\title{PREDICTIONS OF THE \textit{NANCY GRACE ROMAN SPACE TELESCOPE}\\ GALACTIC EXOPLANET SURVEY II:\\ FREE-FLOATING PLANET DETECTION RATES\footnote{During the preparation of this manuscript the name of \textit{The Wide Field Infrared Survey Telescope} was changed to the \textit{Nancy Grace Roman Space Telescope}.}}

\correspondingauthor{Samson A. Johnson}
\email{johnson.7080@osu.edu}

\author[0000-0001-9397-4768]{Samson A. Johnson}
\affiliation{Department of Astronomy, The Ohio State University, 140 West 18th Avenue, Columbus OH 43210, USA}

\author[0000-0001-7506-5640]{Matthew Penny}
\affiliation{Department of Physics and Astronomy, Louisiana State University, Baton Rouge, LA 70803, USA}

\author[0000-0003-0395-9869]{B. Scott Gaudi}
\affiliation{Department of Astronomy, The Ohio State University, 140 West 18th Avenue, Columbus OH 43210, USA}

\author{Eamonn Kerins}
\affiliation{Jodrell Bank Centre for Astrophysics, Alan Turing Building, University of Manchester, Manchester M13 9PL, UK}

\author[0000-0001-5069-319X]{Nicholas J. Rattenbury}
\affiliation{Department of Physics, University of Auckland, Private Bag 92019, Auckland, New Zealand}

\author[0000-0001-8654-9499]{Annie C. Robin}
\affiliation{Institut Utinam, CNRS UMR 6213, OSU THETA, Universite Bourgogne-Franche-Com\'te, 41bis avenue de l’Observatoire, F-25000 Besan\c{c}on, France}

\author[0000-0002-7669-1069]{Sebastiano Calchi Novati}
\affiliation{IPAC, Mail Code 100-22, Caltech, 1200 East California Boulevard, Pasadena, CA 91125, USA}

\author[0000-0002-7669-1069]{Calen B. Henderson}
\affiliation{IPAC, Mail Code 100-22, Caltech, 1200 East California Boulevard, Pasadena, CA 91125, USA}

\begin{abstract}

The \textit{Nancy Grace Roman Space Telescope} (\romanst) will perform a Galactic Exoplanet Survey (RGES) to discover bound exoplanets with semi-major axes greater than 1 au using gravitational microlensing.  
\romanst will even be sensitive to planetary mass objects that are not gravitationally bound to any host star.
Such free-floating planetary mass objects (FFPs) will be detected as isolated microlensing events with timescales shorter than a few days.
A measurement of the abundance and mass function of FFPs is a powerful diagnostic of the formation and evolution of planetary systems, as well as the physics of the formation of isolated objects via direct collapse. 
We show that \romanst will be sensitive to FFP lenses that have masses from that of Mars ($0.1 M_\oplus$) to gas giants ($M\gtrsim100M_\oplus$) as isolated lensing events with timescales from a few hours to several tens of days, respectively. 
We investigate the impact of the detection criteria on the survey, especially in the presence of finite-source effects for low-mass lenses.
The number of detections will depend on the abundance of such FFPs as a function of mass, which is at present poorly constrained. 
Assuming that FFPs follow the fiducial mass function of cold, bound planets adapted from \citet{cassan2012}, we estimate that \romanst will detect $\sim250$ FFPs with masses down to that of Mars (including $\sim 60$ with masses $\le M_\oplus$). 
We also predict that \romanst will improve the upper limits on FFP populations by at least an order of magnitude compared to currently-existing constraints. 
\end{abstract}

\section{Introduction} 
\label{sec:intro}

Time and again, surprising results have arisen from searches for planets beyond our Solar System. Indeed, one of the first planets discovered defined a population of ``hot Jupiters" \citep[e.g.,][]{mayor1995}.  These gas giant planets have orbital periods on the order of days and can have equilibrium temperatures hotter than many stars \citep{collier2010,gaudi2017}.   The {\it Kepler} mission revealed a substantial population of ``super-Earths'' \citep[][]{leger2009}, planets with radii between that of Earth and Neptune; planets which have no analog in our Solar System.
Strange system architectures and planet hosts add even more variety, including planets in tightly packed systems \citep[][]{gillon2017}, planets orbiting both stars of a binary system \citep{doyle2011},  planetary systems orbiting pulsars \citep{wolszczan1992}, and planetary systems orbiting stars at the very bottom of the main sequence \citep{gillon2017}.  
There appears to be almost no physical constraints on where exoplanets may reside.  

Despite this diversity, our statistical census of exoplanets remains substantially incomplete.  
One area of parameter space that has yet to be fully explored is that of planetary-mass objects that are unbound from any host star.
A population of free-floating planetary mass objects (FFPs) in our Galaxy could have two primary sources.
First, such bodies could be formed in relative isolation.
These would essentially be the lowest mass objects assembled through star-formation processes.
Second, such objects could form in a protoplanetary disk initially bound to a host star, and later become liberated from their host.
Regardless of their origin, we will refer to objects with masses comparable to planets that are not bound to any host as FFPs. 

There are several mechanisms that could lead to the formation of isolated low-mass stellar objects \citep[see][and references therein]{luhman2012}.
Stellar cores can be formed at a range of masses through either gravitational or turbulent compression and fragmentation \citep{bonnell2008}. 
Here, the lowest mass cores would result in the lowest mass compact objects; this process may extend down to planetary-mass objects. 
Alternatively, the accretion of gas onto a protostellar core can be truncated, e.g, by being dynamically ejected from their birth clouds by other cores, or by radiation from nearby hot stars that photoevaporate the envelope from around the forming star \citep[e.g.,][]{bate2009}.

Photometric surveys of star forming regions can constrain populations of such low-mass stellar objects \citep[e.g., ][]{gagne2017}.
These surveys are most sensitive to young objects that have not had time to radiate away their thermal energy from formation and thus remain luminous.
In the field, the first thirteen class-defining Y dwarfs were discovered by \citet{cushing2011} and \citet{kirkpatrick2012} using the \textit{Wide-field Infrared Survey Explorer} \citep{wright2010}.
Modelled masses for these objects are of order tens of Jupiter masses ($\tsub{M}{Jup}$).
Volume-limited searches for ultra-cool field dwarfs \citep[e.g.,][]{bardalez2019} constrain these populations, but their low luminosities limit the number of detections and thus the statistical power of these surveys.  Furthermore, these surveys are unlikely to be sensitive to planets with masses substantially smaller than that of Jupiter, regardless of their ages. 

On the other hand, if the dominant reservoir of FFPs is a population of previously bound planets, there is no shortage of methods to liberate them from their hosts.
Planets can be ejected from their systems by the chaotic processes that occur during planet formation~\citep[e.g.,][]{rasio1996}, stripped from their stars by stellar fly-bys~\citep[e.g.,][]{malmberg2011}, or become unbound during the post-main sequence evolution of their hosts~\citep[e.g.,][]{adams2013}.
\citet{hong2018} predict that planet-planet dynamical interactions could also eject lunar-mass satellites of these planets during the encounters. 
It is important to emphasize that objects in the very lowest-mass regime ($<1\tsub{M}{Jup}$) are very difficult to detect by any radiation they emit, even when they are young \citep{spiegel2012}. 

A robust method to detect isolated planetary mass objects is gravitational microlensing \citep{distefano1999}. 
A microlensing event occurs when a massive body (the lens) passes in front of a background star (the source) within roughly one angular Einstein ring radius $\tsub{\theta}{E}$ of the lens,
\begin{equation}
    \tsub{\theta}{E} = \sqrt{\kappa M\tsub{\pi}{rel}}.
\end{equation}
Here, $M$ is the mass of the lensing body, the constant $\kappa=4G(c^2 \textrm{au})^{-1}=8.14$ mas $M_\odot^{-1}$, and the lens-source relative parallax is $\tsub{\pi}{rel} =1 \textrm{au}\left(\tsub{D}{L}^{-1}-\tsub{D}{S}^{-1}\right)$, where $\tsub{D}{L}$ and $\tsub{D}{S}$ are the distances from the observer to the lens and source, respectively.

When the angular separation of the lens and source is comparable to or smaller than $\tsub{\theta}{E}$, the background source is significantly magnified.
The duration of an event is characterized by the microlensing time scale $\tsub{t}{E}=\tsub{\theta}{E}/\tsub{\mu}{rel}$. 
Thus the size of the Einstein ring in combination with the lens-source relative proper motion ($\tsub{\mu}{rel}$) dictates the duration of the event, which can last from a few hours to a few hundred days, depending on the values of the above variables. 
The primary reason why microlensing is a powerful technique to detect FFPs is that it does not rely on the detection of any light from these essentially dark lenses.

While the phenomenology of typical microlensing events (for which $\tsub{\theta}{E}$ is much greater than the angular source size) is well understood, that of microlensing events due to low-mass objects has not been frequently discussed.  
We therefore include a short review of the phenomenology of low-mass microlensing (specifically when the angular source size is larger than $\tsub{\theta}{E}$) in Appendix \ref{sec:phenomenology}.  

One of the pioneering uses of the technique was the search for the then-viable dark matter candidate Massive Compact Halo Objects, or MACHOs. At the time, the typical mass for these candidates for dark matter was unknown, resulting in the need to design a survey that was sensitive to the full range of timescales mentioned above.
The major microlensing collaborations included the Expérience pour la Recherche d'Objets Sombres (EROS; \citealt{renault1997}, the MACHO collaboration \citep{alcock1997}, the Microlensing Observations in Astrophysics Collaboration (MOA-I, \citealt{muraki1999} and the Optical Gravitation Lens Experiment (OGLE-I, \citealt{udalski1992}).  These collaborations set out to detect these MACHOs by monitoring the Large Magellanic Cloud, searching for microlensing events in this high density stellar source environment, with a large cross-section through the dark matter halo.  
Particularly relevant to this discussion, the combined analysis of the MACHO and EROS surveys demonstrated that $\la 25\%$ of the dark halo is made of planetary-mass MACHOs in the mass range between roughly $0.3$ times the mass of Mars and the mass of Jupiter, the first such constraints on the abundance of planetary-mass objects in halo of our Galaxy \citep{alcock1996}.  
See \citet{moniez2010} for a comprehensive history of these efforts. 

Once MACHOs were largely ruled out as a dark matter candidate, microlensing surveys began to focus on lines-of-sight toward the Galactic bulge to constrain Galactic structure~\citep{paczynski1991} and search for bound exoplanets~\citep{mao1991,gould1992}. 
Initially, these surveys lacked the field of view to both find relatively rare microlensing events and monitor them with sufficient cadence to detect the much shorter (and unpredictable) planetary perturbations. Instead, a two-tier system was employed, wherein the survey teams used relatively low-cadence observations to alert follow-up observers of ongoing microlensing events.  The relatively small numbers of ongoing microlensing events could then be monitored at much higher cadence by collaborations with access to a longitudinally-distributed suite of telescopes.
See \citet{gaudi2012} for a review of the history of microlensing surveys for exoplanets during this phase of the field.

Eventually, the MOA and OGLE surveys, along with the (more recently formed) Korea Microlensing Network (KMTNet, \citealt{kim2016}) survey, have developed the capability to monitor the Galactic bulge with sufficient cadence to simultaneously detect isolated microlensing events and search for perturbations due to bound planets. 
This resulted in the first tentative detection of an excess of $\sim 1$~day long events, which implied a substantial population of Jupiter-mass FFPs with an inferred abundance of roughly two free-floating Jupiter-mass planets per star in the Galaxy \citep{sumi2011}. 

This result was later challenged by \citet{mroz2017}, who placed an upper limit of $\la 0.25$ Jupiter-mass FFPs per star.
Notably though, \citet{mroz2017} did find tentative evidence of an excess of very short timescale events ($\tsub{t}{E}\lesssim0.5$ d), possibly indicating a population of free-floating or wide-separation Earth-mass planets, although it is important to note that these events were generally poorly sampled and thus have large uncertainties in their timescales. They therefore may be spurious.
Regardless, these efforts demonstrate the potential of Galactic bulge microlensing surveys to find free-floating or widely-bound planetary-mass objects.  

Indeed, quite recently, multiple well-characterized, extremely-short microlensing events have been discovered. 
\citet{mroz2018a}, \citet{mroz2019a}, and \citet{mroz2020} together report a total of four FFP candidates, two of which had timescales consistent with Earth- or Neptune-mass lenses. 
\citet{han2020b} report the discovery of three events consistent with brown dwarf mass lenses (masses $\sim0.04 M_\odot$), of which two are isolated and one is in a near equal-mass binary. 
An important caveat for candidate FFP events is the potential to exclude of any potential host stars.
If the separation of a planet and its host is sufficiently large $(\gtrsim10$ au; \citealt{han2005}) and the geometry is correct, the source can appear to be magnified by an effectively isolated planet.
Thus, wide-separation planets can masquerade as FFPs in a subset of microlensing events.

This has been discussed before by several authors \citep{distefano1999,han2003,han2005}, all of which propose pathways to determine whether a planetary mass lens is bound or free-floating. 
\citet{mroz2018a} and \citet{mroz2019a} place limits on the presence of a host photometrically, but detailed modelling of the magnification curve and photometric follow-up can also be used to determine if the lens is isolated \citep{han2003,han2005,henderson2016a}.
As an example, detailed modeling has been used to determine the true, bound nature of an FFP candidates by \citet{bennett2012} and \citet{han2020a}.

Keeping in mind these caveats, it has been demonstrated previously \citep[][]{bennett2002,strigari2012,penny2013,ban2016,henderson2016a,penny2017} that a space-based microlensing survey will have unprecedented sensitivity to short-timescale microlensing events due to FFP lenses that have masses comparable to our Moon or greater.  
We investigate this opportunity more fully here, as applied to the NASA's next flagship mission, the \textit{Nancy Grace Roman Space Telescope (Roman)}. 

\subsection{The Nancy Grace Roman Space Telescope and its Galactic Exoplanet Survey} 
\label{sec:primer}

Initially called the \textit{Wide Field Infrared Survey Telescope} \citep[\wfirst,][]{spergel2015}, \romanst is currently planned to conduct three Core Community Surveys: the High Latitude Survey \citep{troxel2019}, the Type Ia Supernovae Survey (photometric \citep{hounsell2018} and spectroscopic), and the Galactic Exoplanet Survey \citep{penny2019}.   
These surveys will be accompanied by a Guest Observer program (including notionally 25\% of observing time) and a demonstration of numerous new-to-space technologies with the Coronagraph Instrument (CGI, \citealt{debes2016,bailey2019}).

The surveys currently have notional designs that will allow them to make key measurements that will in turn provide unique constraints on the nature and time evolution of dark matter and dark energy, as well as provide novel constraints on the demographics of cold exoplanets \citep{akeson2019}.
The designs of these surveys are notional in that the final observing program will not be settled on until much closer to launch, and, importantly, will incorporate community input.

For the Roman Galactic Exoplanet Survey, \romanst will use the microlensing technique to search for bound planets with mass roughly greater than that of Earth ($M_\oplus$) with semi-major axes in the range of $\sim1-10$ Astronomical Units (au)\footnote{\romanst will also discover $\gtrsim$100,000 planets with periods $\lesssim$64 days using the transit technique \citep{montet2017}.}. 
At planet-host star separations roughly equivalent to the Einstein radius of the lens system (and thus peak sensitivity), \romanst will be able to detect planets with masses as low as roughly twice the mass of the Moon, roughly the mass of Ganymede \citep[][hereafter \citetalias{penny2019}]{penny2019}.  
Through finding these planets near and beyond the water snowline of host stars, \romanst will complement the parameter space surveyed by {\it Kepler} \citep{Borucki2010}.
When combined, these broad, monolithic surveys promise to provide the most comprehensive view of exoplanet demographics to date, and thus provide the fundamental empirical data set by which predictions of planet formation theories can be tested \citep{penny2019}.

The current version of the \romanst microlensing survey area covers approximately 2 deg$^2$ near the Galactic bulge, comprised of 7 fields covered by the 0.282 deg$^2$ field-of-view of the Wide Field Instrument \citep[WFI,][]{spergel2015}. 
Throughout the survey, it will observe some $\sim$50,000 microlensing events of which roughly 1400 are predicted to show planetary perturbations \citepalias{penny2019}. 
The current notional survey design includes six 72-day seasons, clustered near the beginning and end of the 5 yr primary lifetime of the mission.
Each season will be centered on either the vernal or autumnal equinoxes, when the Galactic bulge is visible by \romanst.

During a season, \romanst~will perform continual observations using its wide 1--2 $\mu$m \textit{W146} filter at 15 min cadence.
Each visit will have a 46.8 sec \textit{W146} exposure of the WFI that will reach a precision of 0.01 mag at \textit{W146}$\approx$21.
These observations will be supplemented with at least one and likely two narrower filters (yet to be decided), which will sample the fields at much lower cadence.  
\citetalias{penny2019} assumed observations with only one additional (\textit{Z087}) filter with a 12 hr cadence, but this observing sequence has not yet been finalized.
When a microlensing source star is sufficiently magnified and observations are taken in more than one filter, \romanst~will be able to measure the color of the microlensed source star.
Measurement of the source color and magnitude can be used to constrain the angular radius of the source star $\theta_*$, which can be be used to measure $\tsub{\theta}{E}$ if the event exhibits finite-source effects \citep{yoo2004}.
For more details on the currently planned \romanst hardware, the microlensing survey design, and the bound planet yield, the reader is encouraged to read \citetalias{penny2019}.

\subsection{Constraining the Abundance of Free-Floating Planets with \romanst} 
\label{sec:primer2}

The properties of \romanst and the Galactic Exoplanet Survey design that make it superb at detecting and characterizing bound planets are the same properties that allow it to detect and characterize FFPs. 
FFPs can produce events lasting from $\sim$ hr to $\sim$ day.  
Many of the same observables for bound planet microlensing events are also desirable for FFPs, such as the source color and brightness, which can constrain the angular source size, and the mass of the lensing body.
Measuring the mass of an isolated lens requires additional measurements of event parameters \citep{gould1996}, and would require supplementary and simultaneous ground-based or space-based observations \citep[e.g., by \textit{EUCLID},][]{zhu2016,bachelet2019,ban2020}. 
We do not address parameter recovery through modelling, nor mass estimation of detected lenses, both of which are beyond the scope of this work. 

The goal of this work is to predict \romanst's ability to measure the distribution of short-timescale events attributed to free-floating planets.
To do so, we will briefly revisit the microlensing survey simulations presented in \citetalias{penny2019} and detail the changes we made to them in Section \ref{sec:sims}. 
We then examine light curves \romanst will detect in Section \ref{sec:lcs}.
Section \ref{sec:yield} will contain a discussion of the yield and limits \romanst will place on FFPs in the Milky Way.
Finally, we will discuss our findings and conclude in Sections \ref{sec:disc} and \ref{sec:conclusion}. 
We include two appendices, one that provides a primer on the phenomenology of microlensing events in the regime where the angular size of the source is much greater than the angular Einstein ring radius (Appendix \ref{sec:phenomenology}) and a second exploring the sensitivity of \romanst's yield to the detection criteria we impose (Appendix \ref{sec:det_thresh}).

\section{Simulations} 
\label{sec:sims}

To simulate the \romanst{} microlensing survey we use the free-floating planet module of the \gulls microlensing simulator~\citep{penny2013, penny2019}. 
Here we only briefly discuss how FFP simulations differ from the bound planet simulations of \citetalias{penny2019}. 
We use the mission and survey parameters for the Cycle 7 design as fully detailed in \citetalias{penny2019} and summarized in Table \ref{tbl:wfirstsummary}.

\begin{table}
  \caption{\romanst\ Galactic Exoplanet Survey Parameters}
  \begin{tabular}{ll}
    \hline
    Area & 1.97~deg$^2$\\
    Baseline & 4.5 years\\
    Seasons & $6\times 72$~days\\
    Fields & 7\\
    Avg. Slew and Settle & 83.1 s\\
    Primary (\textit{W146}) filter & 0.93-2.00 $\mu$m\\
    \hspace{10pt} exposure time & 46.8 s\\
    \hspace{10pt} cadence & $15$~minutes\\
    \hspace{10pt} total exposures & ${\sim}41,000$ per field\\
    Secondary (\textit{Z087}) filter & 0.76-0.98 $\mu$m\\
    \hspace{10pt} exposure time & 286 s\\
    \hspace{10pt} cadence & $\lesssim12$~hours\\
    \hspace{10pt} total exposures & ${\sim}860$ per field\\
    Phot. Precision & 0.01 mag @ \textit{W146}${\sim}21.15$\\
    \hline
  \end{tabular}
      {\bf Notes}: A summary of the Cycle 7 design is fully detailed in \citetalias{penny2019}. 
      This is the current design, and is subject to change prior to the mission.
      For example, the exposure time and cadence of observations in the \textit{Z087} and other filters has not been set; we have assumed a $12$ hour cadence here, but observations in the other filters are likely to be more frequent.
      \label{tbl:wfirstsummary}
\end{table}

\gulls simulates individual microlensing events by combining pairs of source and lens stars drawn from a population synthesis Galactic model (GM). 
We use the same GM as \citet{penny2013} and \citetalias{penny2019}, version 1106 of the Besan{\c c}on model, for consistency between our results. Version 1106 is intermediate between the two publicly available Besan{\c c}on model \citep{robin2003,robin2012}, and is described fully in \citet{penny2013} and \citetalias{penny2019}.
The usefulness of population synthesis GMs for microlensing was first demonstrated by \citet{kerins2009}. 
An updated model by \citet{specht2020} has recently been shown to provide a high level of agreement with the 8,000-event OGLE-IV event sample of \citet{mroz2019}.

\gulls simulates \romanst's photometric measurements by injecting GM stars, including the source, into a synthetic postage stamp image. From this image the photometric precision as a function of magnification is computed assuming a $3\times3$ pixel square aperture centered on the microlensing event.

The actual \romanst photometric pipeline will be much more sophisticated than this, using both point spread function (PSF) fitting and difference image analysis to perform photometry. 
Aperture photometry is likely somewhat conservative relative to PSF fitting photometry in terms of photon noise, but this is offset by optimism in not dealing with relative pixel phase offsets with an undersampled PSF (see \citetalias{penny2019} for a full discussion).
The model microlensing light curve is computed from a finite-source point lens model \citep{witt1994} with no limb-darkening.  
The realistic, color-dependent redistribution of surface brightness from limb darkening will modify the light curve shape of events in which finite-source effects are present \citep{witt1995, herovsky2003}, but does not significantly affect detection probability. We briefly discuss the impact of omitting limb darkening from our simulations in Section \ref{subsec:limb}. 

Our simulations follow those of \citetalias{penny2019} almost exactly, but we replace the stellar lenses drawn from the catalogs generated from the GM with an isolated planetary mass object and assume zero flux from the injected lens. 
This results in all simulated events having planetary-mass point lenses with the velocity and distance distributions of stars in the GM. 
One might expect small differences in the phase-space distributions between stars and FFPs, depending on their origin, but we do not account for this in this study \citep[e.g.,][found FFPs are ejected from clusters with larger velocities than escaping stars, but only by a few km~s$^{-1}$, which is much less than typical ${\sim}100$~km~s$^{-1}$ relative velocities between lens and source]{vanelteren2019}. 

The source and lens of each simulated microlensing event are drawn from GM catalogs that represent a $0.25\times0.25$~deg$^2$ area of sky, which we call a sight line.
Each event $i$ is assigned a weight $w_i$ proportional to the event's contribution to the total event rate along a sight line,
\begin{equation}
\label{eqn:weight}
w_i = 0.25^2\text{deg}^2\tsub{f}{1106,\romanst}\Gamma_{\text{deg}^2}\tsub{T}{sim}u_{0,\text{max},i}\frac{2\mu_{\text{\tiny rel},i}\theta_{\textrm{\tiny{E}},i}}{W},
\end{equation}
where $\tsub{T}{sim}=6\times 72$~d is the total \romanst microlensing survey duration, $u_{0,\text{max,i}}$ is the maximum impact parameter for each simulated event, $\Gamma_{\text{deg}^2}$ is the sight-line's microlensing event rate per square degree, $\tsub{f}{1106,\romanst}$ is a correction factor, and $W$ is a normalization factor defined below. 
The $\Gamma_{\text{deg}^2}$ event rates were calculated by Monte Carlo integration using catalogs of source and lens stars drawn from the GM.

We use the same $\tsub{f}{1106,\romanst}=2.81$ as in \citetalias{penny2019} which matches the GM's event rate to the microlensing event rate measured using red clump source stars by \citet{Sumi2013} and corrected by \citet{Sumi2016}.
\citet{mroz2019} measured microlensing event rates with a larger sample of events from the OGLE-IV survey. They measured the event rate per star for source stars brighter than $I<21$ (a so-called all-star event rate) that was consistent with the \citet{Sumi2016} red-clump event rate, but which was a factor of 1.4 smaller than MOA's all star event rate estimated by \citet{Sumi2016} with sources brighter than $I<20$. We elect to maintain the same event rate scaling as \citetalias{penny2019} because the origin of the discrepancy in all-star rates between \citet{mroz2019} and \citet{Sumi2016} is not clear, and for reasons discussed in \citetalias{penny2019} we expect that the small bar angle in the GM may cause an over-correction if corrections are tied to all-star event rates.

For each event, $u_0$ is uniformly drawn from [0,$u_{0,\text{max}}$], where $u_{0,\text{max},i}=\max({1,2\rho})$ and $\rho=\theta_*/\tsub{\theta}{E}$ is the angular radius of the source star relative to $\tsub{\theta}{E}$.
We impose the $2\rho$ alternative to ensure that all lens transiting source events are simulated.
We also ran supplemental simulations at higher masses with $u_{0,\text{max},i}=\max({3,2\rho})$ and found consistent event rates to those use $u_{0,\text{max},i}=\max({1,2\rho})$.
This event weight should be normalized to the stellar-lens event rate, so we divide $\theta_{\textrm{\tiny{E}},i}$ by the mass ratio of the injected lens ($M_{p,i}$) and the star that it is replacing ($M_{*,i}$), $\sqrt{q_i}=\sqrt{M_{p,i}/M_{*,i}}$, to correct the value.
We note that this methodology is equivalent to the assumption that there is one FFP per star in the Galaxy.

The normalization factor is then defined 
\begin{equation}
W=\sum_i\frac{2\mu_{\text{\tiny rel},i}\theta_{\textrm{\tiny{E}},i}}{\sqrt{q_i}}.
\end{equation}
such that the sum of all simulated event weights would equal the number of events occurring over the survey duration had each stellar lens not been replaced by an FFP.

We run two sets of simulations, both of which have lens masses drawn from  $\log(M/M_\oplus)\in[-5,5]$ (i.e., 0.5\% the mass of Pluto to 0.3 $M_\odot$). 
In the first set of simulations, we simulate equal numbers of planets with a range of discrete masses uniformly spaced by 0.25 dex. 
In the second set, we draw log-uniform random-mass lenses from the same range. 
In both cases we draw events until the error on the estimated event rate due to random sampling and accounting for unequal event rates is less than 0.1\%. 

\subsection{Detection Criteria}

We use two detection criteria for microlensing events. The first is the difference in $\chi^2$ of the observed lightcurve relative to a flat (unvarying) light curve fit
\begin{equation}
\Delta\chi^2= \tsub{\chi}{Line}^2 -\tsub{\chi}{FSPL}^2,
\end{equation}
where $\tsub{\chi}{Line}^2$ is the $\chi^2$ value of the simulated light curve data for a flat line at the baseline flux and $\tsub{\chi}{FSPL}^2$ is the same but for the simulated data to the true finite-source point lens model of the event.

The second criteria is that $n_{3\sigma}$, the number of consecutive data points measured at least $3\sigma$ above the baseline flux, must be greater than 6, i.e.,
\begin{equation}
    n_{3\sigma}\geq6.
\end{equation}
This criteria serves two purposes. 
First, it mimics the type of selection cut that previous free-floating planet searches have used to minimize the number of false-positives caused by multiple consecutive outliers from long-tailed uncertainty distributions \citep[e.g.,][]{sumi2011,mroz2017}. 
Second, it ensures that any events detected will stand a good chance of being modeled with 4 or 5 free parameters without over fitting. 
Extremely short events and those with large $\rho$ may suffer from degeneracies where even six data points may be insufficient to correctly model the event (Johnson et al., in prep.).
Naively, the probability of six consecutive data points (assuming that they are Gaussian distributed) randomly passing this criterion is $\sim (1-0.9987)^6\approx10^{-18}$.
Given the number of data points per light curve ($\sim4\times10^5$) and the number of light curves ($\sim2\times10^8$), we expect a by-chance run of 6 points more than $3\sigma$ to occur at a rate of $\sim4\times10^{-4}$ per survey.
Thus this may appear to be an overly conservative detection criterion. 
However, it is likely that neighboring data points will not be strictly uncorrelated, and therefore it is important to adopt conservative detection criteria.  
For example, a false positive event may be caused by hot pixels, which are obviously correlated with time.  
We briefly discuss the possibility of contamination by this and other false positives in Section \ref{subsec:false_pos}.
We further motivate these selection criteria in the next section.

Our predictions for the yields of detectable free-floating planets are calculated using the weights defined in Equation \ref{eqn:weight} modified by a Heaviside step function for each detection criteria,
\begin{equation}
\label{eqn:ndet}
\tsub{N}{det}=\sum_i{w_i H(\Delta\chi^2-300) H(n_{3\sigma}-6)}.
\end{equation}

\section{Lightcurves of Free-Floating Planets as seen by Roman}
\label{sec:lcs}

The continual coverage provided by \romanst enables the detection of the microlensing events caused by free-floating planets without the difficulties faced by ground-based microlensing surveys. 
In this section we explore the light curves of free-floating planets that \romanst might detect, covering a wide range of planet masses.

\begin{figure}[t]
\epsscale{\epsScaleFactor}
\plotone{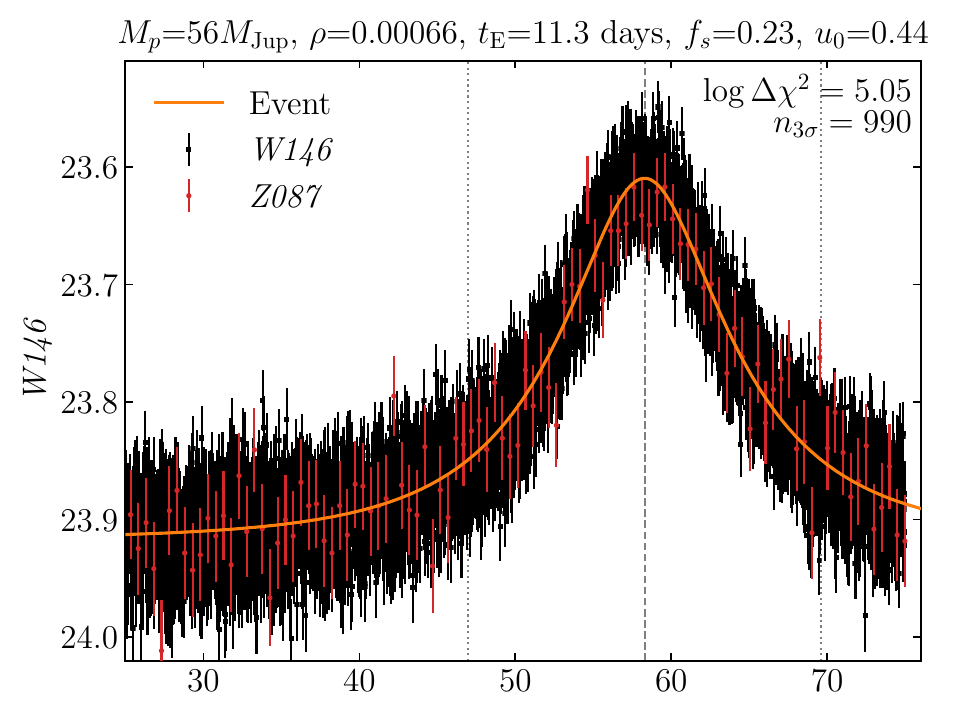}
\plotone{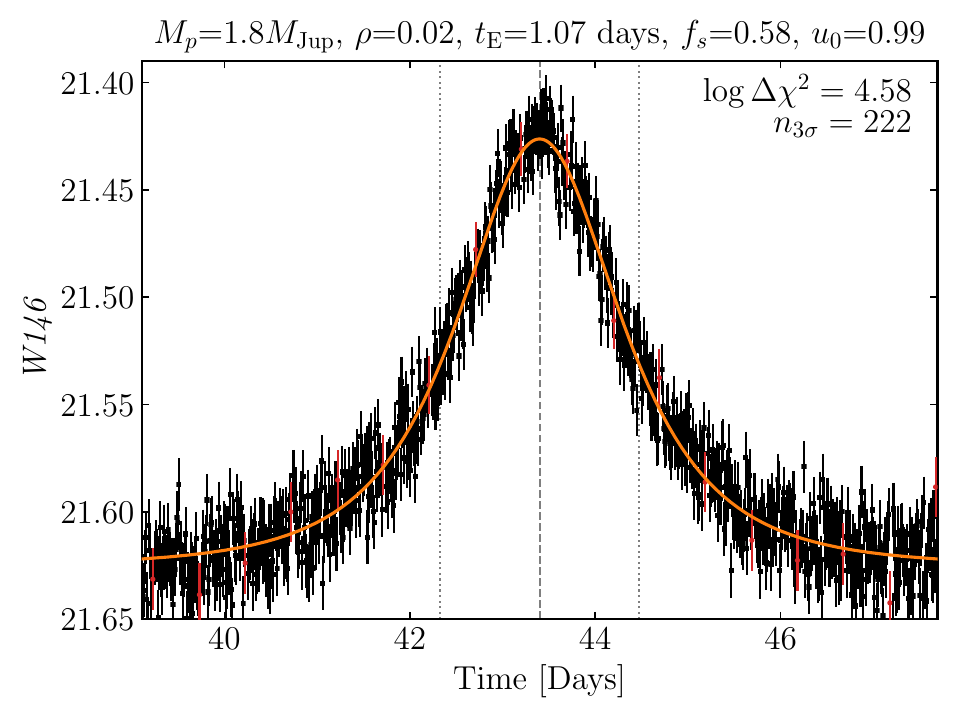}
\caption{
Two examples of simulated events as observed by \romanst.
Black (red) points are observations in the \textit{W146} (\textit{Z087}) filters, and the overlying orange line is the input lensing model.
Above each panel, $M_p$ is the mass of the lens in Jupiter masses ($\tsub{M}{Jup}$) or Earth masses ($M_\oplus$), $\rho$ is the angular size of the source normalized to the Einstein ring, $\tsub{t}{E}$ is the Einstein timescale of the event, $f_s$ is the blending parameter, and $u_0=\theta_0/\tsub{\theta}{E}$ is the minimum impact parameter. 
We also include the values of $\log{\Delta\chi^2}$ and $n_{3\sigma}$ light curve.
Vertical short-dashed gray lines indicate $\pm\tsub{t}{E}$ values of the event, and the long-dashed grey line the peak of the event.
The expected photometric precision and 15 min cadence for observations in the primary \textit{W146} band will make detection of such events trivial. 
\textit{Upper left:} An event with a $\sim 60~M_{\rm Jup}$ brown dwarf lens. 
\textit{Upper right:} An event with a $\sim 2~M_{\rm Jup}$ mass lens. 
\label{fig:lc_1}}
\end{figure}

\begin{figure}[t]
\epsscale{\epsScaleFactor}
\plotone{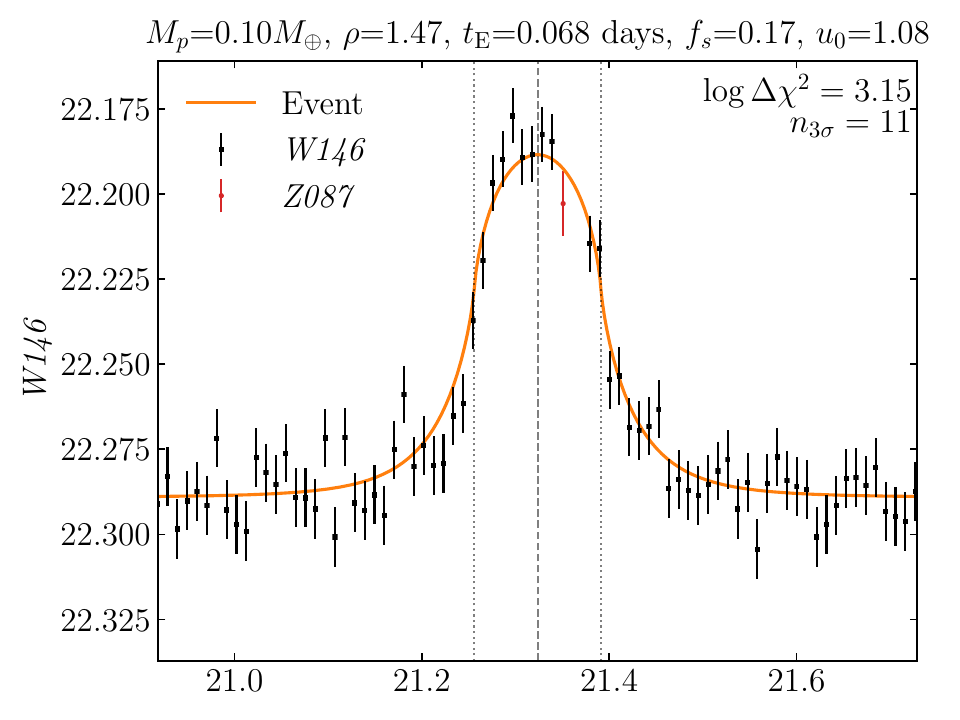}
\plotone{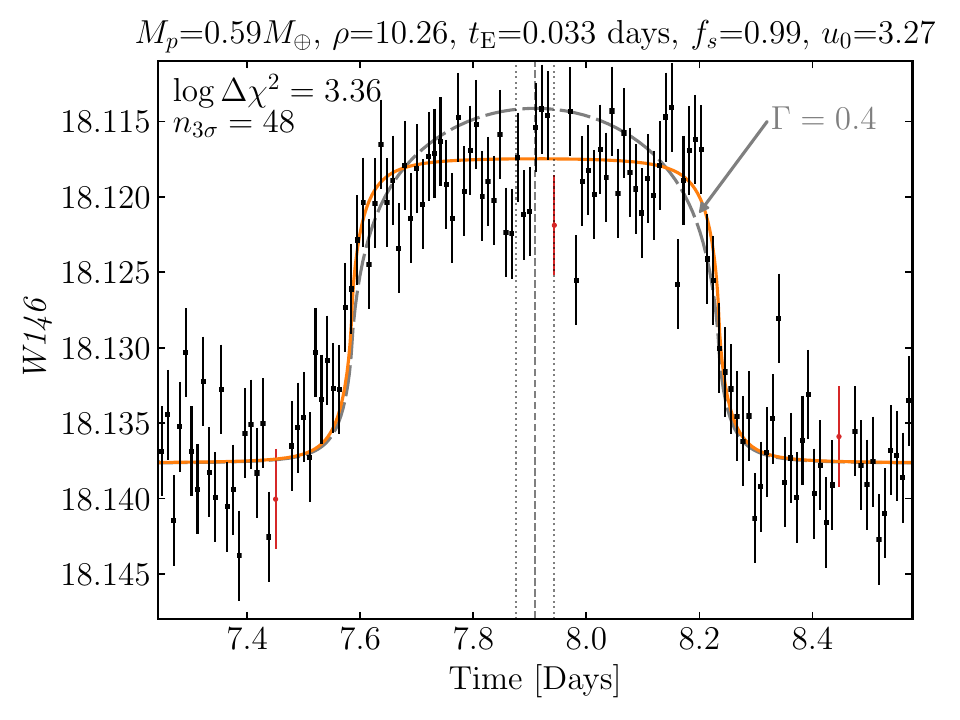}
\caption{
Same as Figure \ref{fig:lc_1}, but for two very low-mass lenses.
We note that, although both events contain one measurement taken in the \textit{Z087} filter, this is not representative of most low-mass lens events.
\textit{Upper:} Illustrative light curve due to a roughly Mars-mass FFP, with relatively mild finite-source effects. 
\textit{Lower:} Illustrative light curve due to a $\sim 0.6~M_\oplus$ FFP, in this case lensing a giant source, thereby exhibiting strong finite-source effects.
Note that, in this case the fact that the source is a giant results in nearly no blending and the large value of $\rho$.
In such cases, the magnification would saturate at $1+\frac{2}{\rho^2}$ shown as the orange line in the absence of limb darkening. However, when we include limb darkening the lightcurve would appear as long-dashed gray line (for $\Gamma=0.4$)
\label{fig:lc_2}}
\end{figure}

We begin with large-mass FFPs and brown dwarfs, which can be challenging to observe from the ground due to their event timescales being comparable to several days.
Figure~\ref{fig:lc_1} shows the light curve for a brown-dwarf-mass lens in the upper panel.
This example has a relatively long timescale compared to what is expected for typical free-floating planet events, but we include it as an extreme case to demonstrate the confusion with stellar lens events. 
These cases display the density of \romanst~photometry, especially in the lower panel, which has nearly 1000 $3\sigma$-significant \textit{W146} measurements in the time span of roughly 6 days. 
These events will be extremely well characterized, and are nearly guaranteed to have color measurements while the source is magnified. 

Figure \ref{fig:lc_2} show the light curves of events at the opposite end of the detectable FFP mass spectrum. 
A very low-mass lens exhibits modest finite-source effects in the upper panel.
Much stronger finite-source effects are apparent in the lower panel for a giant source with $\rho\approx10$.
In the latter case, the magnification saturates at the expected value of $1+2/\rho^2$ (Equation \ref{eqn:fse_peak}), i.e., just $1.02$  in the absence of limb darkening.
To demonstrate the impact of limb darkening for this event, we use the same event parameters to recompute the magnification using the \citet{lee2009} method as implemented in \texttt{MulensModel} \citep{poleski2018}. 
This is shown as the gray, long-dashed line underlying the simulated event. 
The peak is higher than in the event without limb darkening, and the ``shoulders'' of the top hat drop modestly. 
Even for such extreme finite-source events the impact of limb darkening will be modest on the number of events that pass selection cuts.
Both events highlight the precision of \romanst photometry.
The light curves in Figures \ref{fig:lc_1} and \ref{fig:lc_2} are chosen to demonstrate a number of morphologies, photometric precisions, masses, and detection significances of \romanst events.

\begin{figure*}
\epsscale{\epsScaleFactor}
\plotone{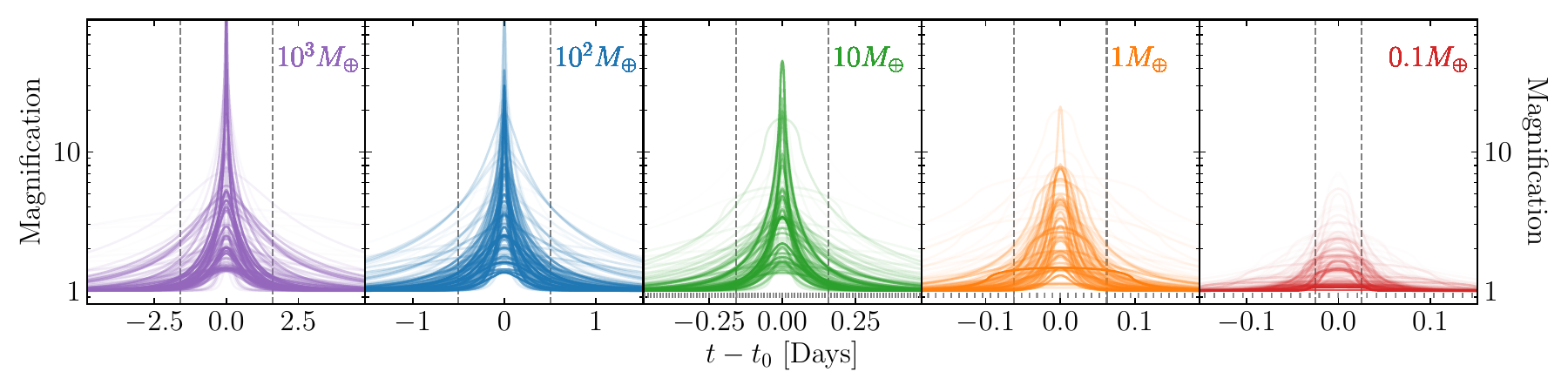}
\caption{
Samples of simulated magnification curves from events detectable by \romanst\ at each mass of $10^3, 10^2, 10, 1, 0.1 M_\oplus$, from left to right.
For each mass, we randomly select 100 events that passed our detection criteria and plot their magnification curves. 
The weighted average $\pm\tsub{t}{E}$ is indicated by the vertical dashed lines in each panel.
Note the horizontal axis scale changes as mass decreases and the vertical axis uses a logarithmic scale.
The black, horizontal tick marks below the curves indicate the \textit{W146} cadence; we note that they are only shown for masses of $10~M_\oplus$ and below. 
The transparency of each curve is proportional to the weight of the event normalized to the maximum weight of events included in the panel.
In this way, darker lines exemplify events that will contribute more to the event rate for that mass bin.
\label{fig:many_lc}}
\end{figure*}

\begin{figure}
\epsscale{\epsScaleFactor}
\plotone{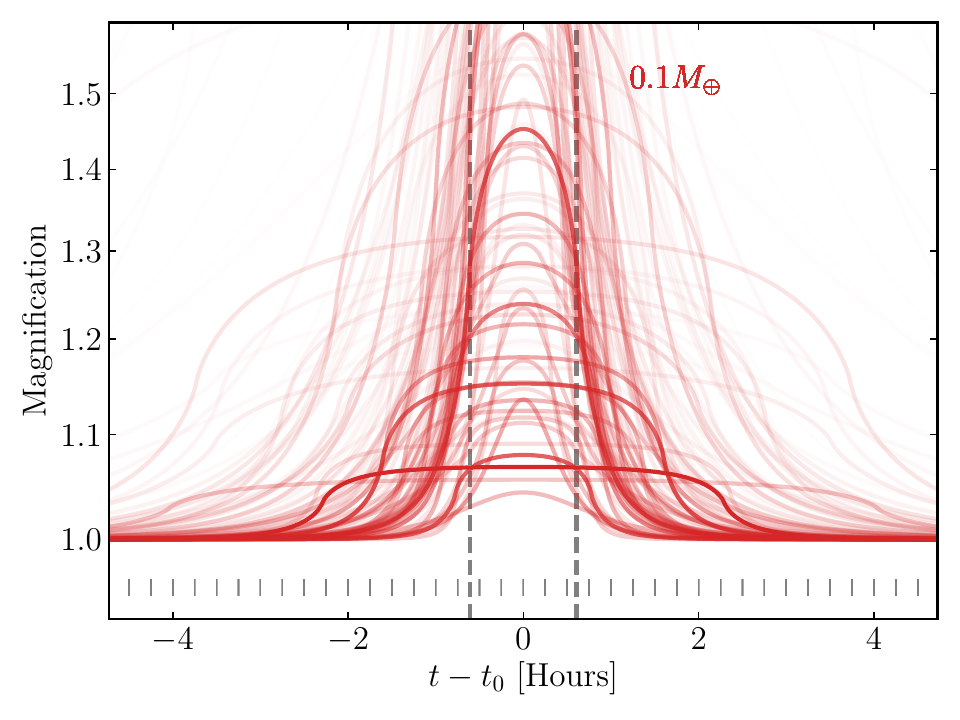}
\caption{The rightmost panel of Figure \ref{fig:many_lc}, but rescaled to highlight the finer detail of light curves arising from $\sim0.1~M_\oplus$ lenses.
Note that the magnification remains in log-scale,
but the horizontal axis has been converted from days to hours.
The vertical dashed lines are the weighted average $\tsub{t}{E}$, which indicate that the Einstein timescales are generally much shorter than the observed timescales, which are set by the crossing time of the source when $\rho \gg 1$.
The gray vertical dashes match the 15 min observing cadence of the \textit{W146} band.
\label{fig:many_01mearth_lc}}
\end{figure}

\begin{figure}
\epsscale{\epsScaleFactor}
\plotone{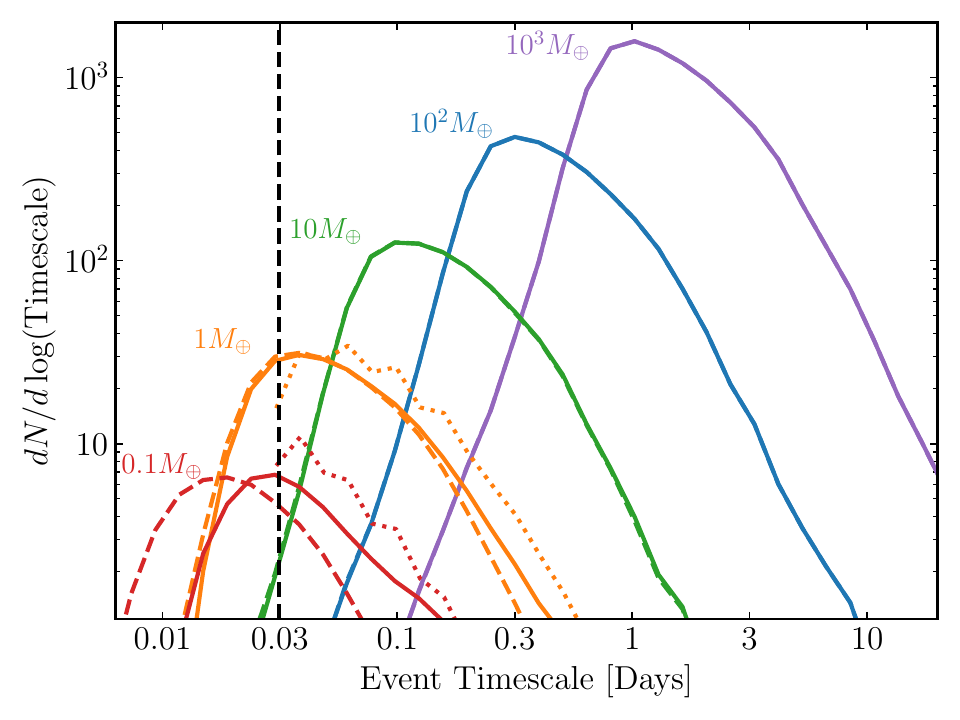}
\caption{
The distribution of detected events as a function of timescale for different lens mass populations. 
We plot the distributions as a function of $\tsub{t}{E}$ as dashed lines. 
The solid lines are the distributions as a function of the maximum of $\tsub{t}{E}$ or the source half-chord crossing time, $0.5t_c$. 
These distributions are nearly identical for masses above $10M_\oplus$ because for these masses typically $\tsub{t}{E}\gg 0.5t_c$, whereas for lower masses the timescale is largely set by the source chord crossing time.
For the two lowest masses, we also plot as dotted lines for the distribution of $\frac{1}{2}n_{3\sigma}\times15$min. The vertical dashed line indicates $3\times$ the \textit{W146} band cadence.  
The cut we impose on $n_{3\sigma}$ (e.g., the dashed vertical line) eliminates events that are formally `significant' according to the $\Delta\chi^2$ criterion, but would likely be poorly characterized due to the small number of significant points.  
Interestingly, as a result of the fact that the effective event timescale saturates at the source chord crossing time for low-mass lenses, many  events pass our cuts that would not in the absence of finite-source effects.
\label{fig:yield_te}}
\end{figure}

For a broader, more representative look at the events \romanst will detect, Figure \ref{fig:many_lc} displays an ensemble of light curves for each of the five discrete mass lenses we consider. 
In each panel, we randomly select 100 events that passed our detection criteria in $\Delta\chi^2$ and $n_{3\sigma}$.
We then normalize the transparency of each curve to the maximum weight of those events included (Equation \ref{eqn:weight}).
In this way, darker curves indicate events that contribute more to the calculated event rate.
We place vertical dashed lines at the positive/negative weighted average of $\tsub{t}{E}$ for these subsets, as well as a horizontal row of gray dashes below the curves representing the \textit{W146} cadence (15 min) in the three rightmost panels.
Note that the scales of the horizontal axes shrink with decreasing mass (as $\tsub{t}{E}\propto M^{1/2}$), but we maintain the scale between the two rightmost panels.
At higher masses ($\geq10^2M_\oplus$) the light curves look like one would expect for point-like sources.
As the mass of the lens decreases, a larger fraction of detected events exhibit finite-source effects as described in Appendix \ref{sec:phenomenology}.

Figure \ref{fig:many_01mearth_lc} shows the rightmost panel of \ref{fig:many_lc} with both axes re-scaled in order to show finer detail for the lowest-mass lenses.  However, note the magnification axis remains logarithmic.
Note there are only 5 dashes (5 photometric measurements) during the expected duration ($2t_{\rm E}$), marked by the vertical gray dashed lines.
However, the true duration of these events is often considerably longer. 
Were there no finite-source effects, the events of low-mass lenses would often be too short to accurately model with the 15 min cadence of the \textit{W146} band, but because the source crossing time for these sources can be a factor of several times longer than $2t_{\rm E}$, these events may be well characterized.

\subsection{Detection Thresholds}

Given the potential challenges involved in detecting short events, we revisit the detection criteria we presented in Section 2 to ensure they fulfill their purpose.
We require that $\Delta\chi^2$ of an event be at least 300 and that the event has an $n_{3\sigma}$ of at least 6. 
These thresholds are similar in nature to the initial cuts placed by \citet{sumi2011} and \citet{mroz2017}.
Both use $n_{3\sigma}\geq3$ as well as a statistic $\chi_{3+}=\sum_i(F_i-\tsub{F}{base})/\sigma_i$ to quantify the significance of candidate events, where $F_i$ is the $i$th data point within an event with uncertainty $\sigma_i$ and $\tsub{F}{base}$ is the baseline flux.
\citet{sumi2011} use $\chi_{3+}\geq80$ while \citet{mroz2017} relaxed this to $\chi_{3+}\geq32$ due to the typically higher quality of the OGLE data.

It is not straight forward to compare our cuts to the $\chi_{3+}$ criteria of \citet{sumi2011} and \citet{mroz2017}, however we can consider an extreme case.
Imagine an event that barely passes both our criteria with $\Delta\chi^2=300$ and $n_{3\sigma}=6$, but with a minimal $\chi_{3+}$.
This event would have 6 consecutive data points, five at $3\sigma$ and a single data point at $16\sigma$, making a total $\Delta\chi^2=301$.
This particular event would then have a value of $\chi_{3+} = 31$, barely failing to pass the \citet{mroz2017} $\chi_{3+}$ threshold.
More realistic events would have higher values of $\chi_{3+}$, therefore our cuts are at least comparable to those used in \citet{mroz2017} but are likely slightly more stringent.
We also expect fewer systematics and less correlated noise in \romanst data compared to that of ground-based surveys.

\citet{sumi2011} and \citet{mroz2017} follow their initial cuts with several more to further vet their samples, ensuring each is truly a microlensing event. 
These include (among others) the rejection of light curves with more than brightening event, the rejection of light curves with poor goodness-of-fit statistics to initial models, and the rejection of events that did not have the rise or fall of the event sufficiently sampled.
Without a detailed investigation of the uncertainties in the observables, we must use heuristic cuts to approximate these detailed investigations. 
We have not implemented these further cuts because our simulations do not contain the false positives they are designed to reject.
We do explore the thresholds we place in Appendix \ref{sec:det_thresh}, and determine scaling relations to predict how loosening or tightening these thresholds will impact \romanst's free-floating planet yield.
These relations can also be used to estimate the change in yield as the microlensing survey design evolves.

We examine how our thresholds of $\Delta\chi^2\geq300$ and $n_{3\sigma}\geq6$ impact the timescale distribution of events in Figure \ref{fig:yield_te}, where we assume delta functions in mass (one planet per star) for each mass shown. 
First, we plot the distribution of events as a function of \mbox{$t=\max(\tsub{t}{E},\frac{1}{2}t_c)$}  as solid lines. 
Here $t_c$ is the source chord crossing time as defined in Equation \ref{eqn:t_c}\footnote{Typically the source radius crossing time as defined by $t_*=\rho\tsub{t}{E}=\theta_*/\tsub{\mu}{rel}$ is used as a proxy for the timescale of the event (e.g., \citealt{skowron2011}), however, we account for non-zero impact parameter $u_{0,*}$ similar to \citet{mroz2019a}. We follow their definition except we use the variable $t_c$ instead for their Equation (10). See Appendix \ref{sec:phenomenology}.}. 

These distributions are meant to show duration of events detected. 
Events that exhibit extreme finite-source effects (and thus have `top hat' light curves) \mbox{$\tsub{t}{E}$ will be less than $\frac{1}{2}t_c$} and the event will be longer than expected. 
This will allow for the detection of events that would not be typically detectable were there no finite-source effects.

Second, we plot the distribution as a function of solely the \tE~values of events as dashed lines.
There is essentially no difference between these distributions for lens masses $\geq 10M_\oplus$, but we see a strong offset between the solid- and dashed-line distributions for the $0.1M_\oplus$ events.
For low-mass lenses, this demonstrates the previous point that some detected events would have expected timescales much shorter than would be detectable considering our requirement on $n_{3\sigma}$. 

Finally, for the two lowest masses we show the distribution as a function of $t_{3\sigma}=\frac{1}{2}n_{3\sigma}\times15$min, half the length of the event while significantly magnified. 
These distributions have no events less than 45 min (the vertical, black dashed line), which is indicative of our detection criteria on $n_{3\sigma}$.
For the 0.1 $M_\oplus$ events, the event timescale saturates at the source chord-crossing timescale for many events, pushing the distribution towards longer durations. 
This is even more enhanced when considering the distribution while significantly magnified. 

\begin{figure}
\epsscale{\epsScaleFactor}
\plotone{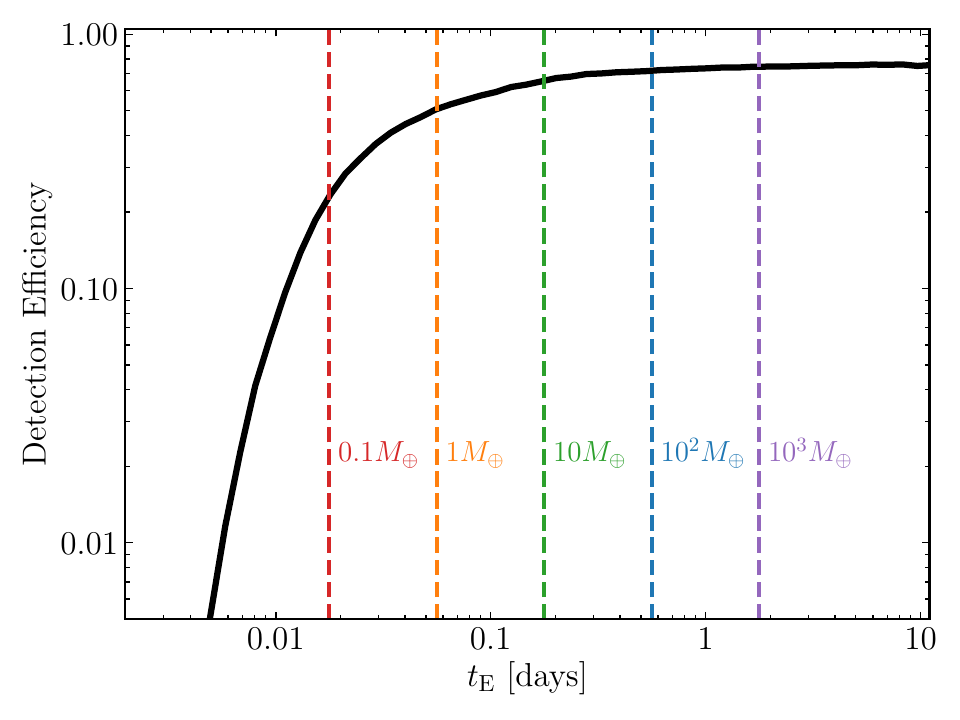}
\caption{
\romanst's detection efficiency as a function of timescale (the black solid line) computed as the fraction of the events that pass our detection criteria relative to all events. 
\romanst will have $>50\%$ detection efficiency down to events with timescales as short as 1.5 hr.
The five vertical lines indicate typical timescales for lenses with the mass indicated.
\label{fig:det_eff}}
\end{figure}

More broadly, we show the detection efficiency as a function of the microlensing timescale $\tsub{t}{E}$ in Figure \ref{fig:det_eff}.
The black line is the number of detected events relative to the number of injected events with a given timescale within a single 72 day season. 
The typical timescales for lenses of five different masses are illustrated with vertical lines. 
Within a season, \romanst will maintain a $\gtrsim50\%$ efficiency down to $\tsub{t}{E}\approx$1.5 hr.
This efficiency would be proportionately lower if we consider the efficiency over the entire 5 yr baseline by a factor of $(6\times72\textrm{ d })/(5\times365 \textrm{ d })=0.23$ if $t_0$ is uniformly distributed, since the Galactic bulge will only be observed for a fraction of a year.

Overall, in this section we have demonstrated that \romanst will be able to detect a wide range and variety of short timescale microlensing events.
This will impact the overall timescale distributions of microlensing events that \romanst will detect, and must need to be accounted for in determining the detection sensitivity used to infer the true underlying distribution of event timescales, regardless of the nature of the lenses.

In the next section, we now present our predictions for the yield of and limits on free-floating planets given the fiducial Cycle 7 survey design. 

\section{Predicted Yields and Limits}\label{sec:yield}

In this section, we present our predictions for the number of FFPs \romanst\ will detect, as well as the limits on the total mass of FFPs that can be set by \romanst.
Recall that the yields are calculated from summing the weights of simulated events that pass our detection cuts using Equation \ref{eqn:ndet}. 
We maintain our detection criteria of $\Delta\chi^2\geq300$ and $n_{3\sigma}\geq6$, but discuss the impact of changing these in Appendix~\ref{sec:det_thresh}.
\begin{figure*}[!ht]
\epsscale{\epsScaleFactor}
\plotone{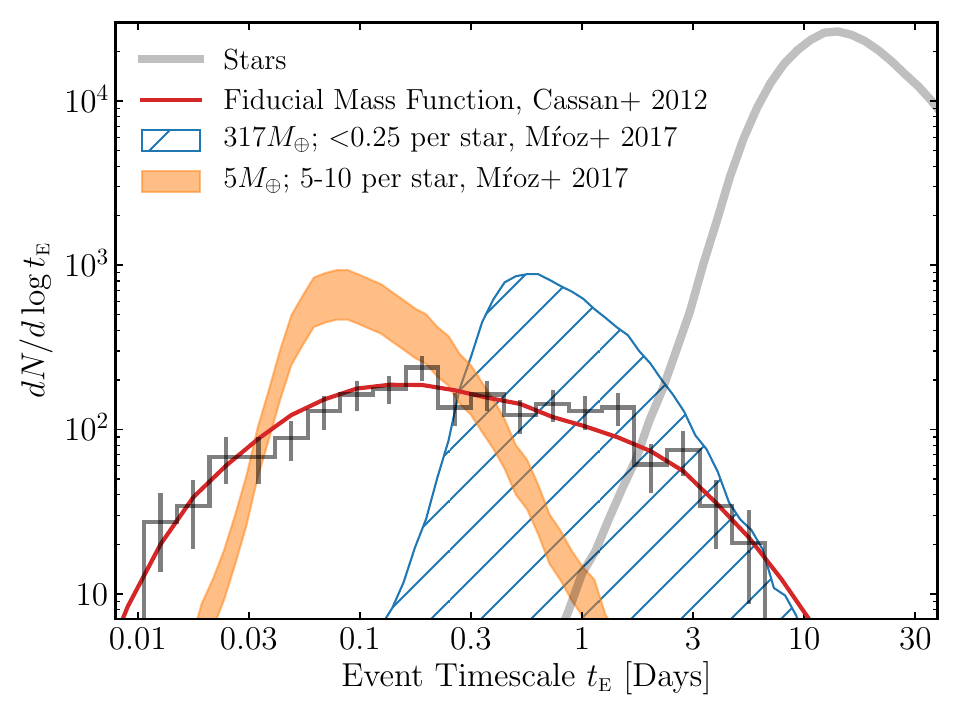}
\caption{
\romanst will be able to measure the timescale distribution arising from FFP over three orders of magnitude in mass.
The expected $\tsub{t}{E}$-signal from the stars is shown as a gray heavy line. 
The blue hatched region marks the upper limit of 0.25 FFPs with mass $\sim \tsub{M}{Jup}$ placed by \citet{mroz2017}.
The orange region marks the expected number of events detected due to $5M_\oplus$ mass lenses assuming there were 5 (lower-bound) to 10 (upper-bound) such FFPs per star in the Galaxy, as tentatively inferred by \citet{mroz2017}.
The red line indicates the timescale distribution of detected FFPs using a fiducial mass function.
The light grey histogram with error bars is a realization of the simulated yield arising from that mass function, including Poisson uncertainties.
\label{fig:te_dist}}
\end{figure*}

\subsection{Yield}
\label{subsec:yield}
We must assume a mass function for FFPs if we are to estimate the number of FFPs that \romanst~will find.  
We assume two forms of mass function, one log-uniform in mass,
\begin{equation}
    \frac{dN}{d \log M_p} = 1~\textrm{dex}^{-1},
\end{equation}
and another inspired by an inferrred mass function of bound planets detected by microlensing \citep[following][]{cassan2012}.
In the second case, we assume that for low-mass planets the mass function saturates at two planets per star below $5.2M_\oplus$.
This prevents the number from `blowing up' as the planet mass decreases. 
The functional form is then
\begin{equation}
\begin{array}{ll}
\frac{dN}{d\log M_p}
=\left\{
\begin{array}{ll}
     \frac{0.24}{\textrm{dex}} \left(\frac{M_p}{95M_\oplus}\right)^{-0.73} & \frac{M_p}{M_\oplus}\geqslant5.2 \\
     2\textrm{ dex}^{-1} & \frac{M_p}{M_\oplus}<5.2
\end{array}
\right.
\end{array}.
\label{eqn:cassan}
\end{equation}
Note that \citetalias{penny2019} used a function with the same mass dependence as the fiducial mass function for bound planets. 
This fiducial function is also consistent with the upper limits on the abundance of bound and wide-orbit planets measured by \cite{mroz2017}. 

This mass function is somewhat optimistic compared to that found by \citet{suzuki2016}, e.g., who found that the mass ratio function is shallower for objects with mass ratio less than $\sim 2\times 10^{-4}$ than that found by \citet{cassan2012}.  This mass ratio corresponds to the typical mass ratio for a Neptune-mass planet.  Nevertheless, we adopt the \cite{cassan2012} mass function for continuity with \citetalias{penny2019}.

We report the expected number of detections as a function of mass in Table \ref{tbl:yield}. 
The first column (`One-Per-Star') assumes that there is a delta function of FFPs at that mass such that there is an equal number of FFPs to stars in the MW. 
The `Log-Uniform' and 'Fiducial' columns assume bins that are 0.5 dex in width for the two mass functions defined above. 
We use the trapezoidal rule to integrate the number of detections with masses from $0.1-1000M_\oplus$ to estimate the total yield of FFPs. 
We include rows for FFPs with masses of 0.01 and $10^4 M_\oplus$ for reference. 
Were the mass function simply log-uniform, nearly 1000 free-floating planets would be detected. 
In the case of the fiducial mass function, we predict that \romanst will detect roughly 250 FFPs. 

Next, we consider how these populations will manifest in the timescale distribution of microlensing events measured by \romanst.
To start, we show the expected timescale distribution of detected stellar events with the same detection criteria ($\Delta\chi^2\geq300, n_{3\sigma}\geq6$) in Figure \ref{fig:te_dist}.
Note that the minimum mass included in the Galactic Model is $0.08 M_\odot\approx80\tsub{M}{Jup}$ in the Galactic disk and $0.15 M_\odot$ in the Galactic bulge.
Then we consider three cases for populations of FFPs.
The blue hatched region has an upper boundary that reflects the limit of at most 0.25 Jovian planets per star from \citet{mroz2017}.
We also include consider the population of $5$-$M_\oplus$ free-floating or wide-separation planets that \citet{mroz2017} cautiously consider as a possible explanation of the excess of very short-timescale events. 
The orange shaded region has a lower (upper) bound corresponding to 5 (10) FFPs per star in the MW that are $5$-$M_\oplus$.
Thirdly, we show the expected distribution of detections using the continuous fiducial mass function in red.
We also draw a realization of this mass function which is included as the gray histogram with Poisson error bars.

If our fiducial assumptions are reasonable, \romanst will be able to detect the signature of terrestrial mass to Jovian mass lenses in the event timescale distribution. 
With the lowest mass planets giving rise to events with extended timescales due to finite-source effects, the sensitivity is pushed to events with lens masses as low as a few times that of Mars. 
The fiducial mass function we use produces events detectable by \romanst with timescales stretching over three orders of magnitude.
These will leak into the timescale distribution attributable to the stars in the Galaxy, but because the model truncates at 0.08 $M_\odot$ there is no smooth transition.

\begin{deluxetable}{lrrr}
 \tablecolumns{4} 
\tablecaption{Expected Free-Floating Planet Yields\label{tbl:yield}}
\tablehead{
\colhead{Mass} & \multicolumn{3}{c}{Mass Function}\\
\colhead{($M_\oplus$)} & \colhead{One-Per-Star} &\colhead{Log-Uniform} & \colhead{Fiducial}
}
\startdata
0.01 & $1.22$ & $0.349$ & $0.698$ \\
0.1 & $17.9$ & $5.13$ & $10.3$ \\
1 & $88.3$ & $25.2$ & $50.5$ \\
10 & $349$ & $83.0$ & $103.$ \\
100 & $1250$ & $298$ & $68.9$ \\
1000 & $4100$ & $976$ & $42.0$ \\
10000 & $13300$ & $3170$ & $25.4$ \\
\hline
Total & $3750$ & $897$ & $249$ \\

\enddata
\tablecomments{The `Total' row is an integration using the trapezoidal rule from $0.1-1000 M_\oplus$. The first and last rows are included for reference.}
\end{deluxetable}

\subsection{Limits}
\label{subsec:limits}

\begin{figure*}[t]
\plotone{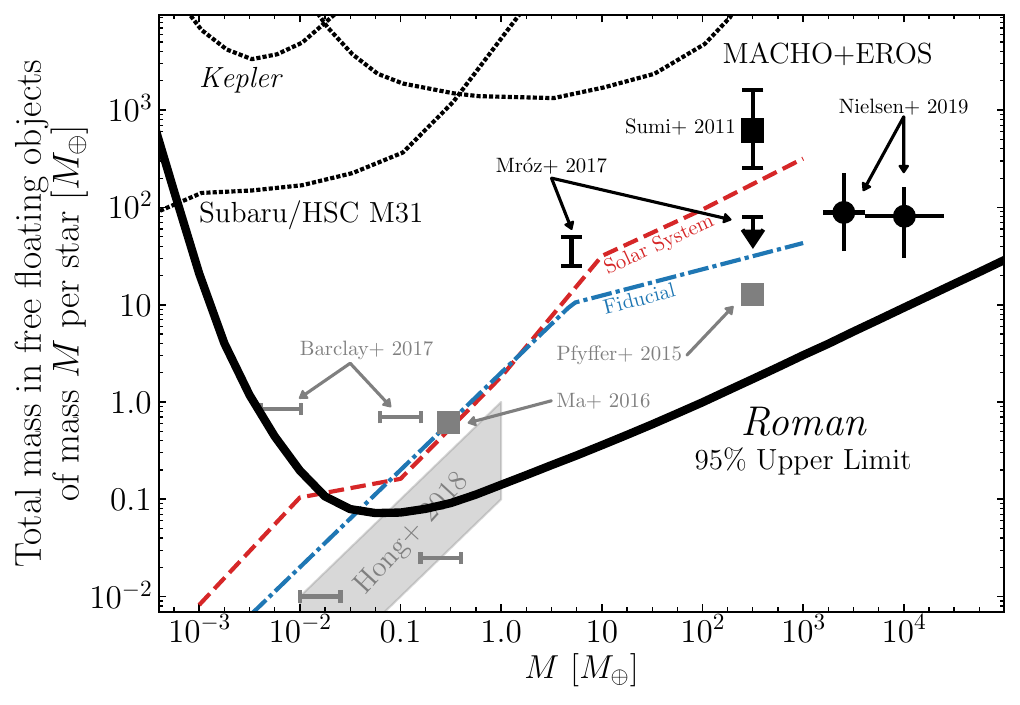}
\caption{The heavy solid black line shows the 95\% confidence upper limit on the total mass of objects per star as a function of the object mass that \romanst~will be able to place if no objects of a fixed mass are detected. It is orders of magnitudes lower than past limits and can test predictions on the abundance of FFPs from planetary formation (or free floating compact objects formed from other mechanisms, such as primordial black holes).
The black dashed lines represent similar limits placed by microlensing searches for massive compact halo objects.
The blue dot-dashed line shows our fiducial mass function (Equation \ref{eqn:cassan}).
For context, the red dashed line shows the case if roughly a Solar System's worth of objects per star were free floating in the Galaxy.
The observational results of previous microlensing surveys are plotted in black points indicated by `Sumi+ 2011' and `M\'{r}oz+ 2017'.
The black circles are frequencies for widely separated bound planets reported by \citet{nielsen2019} using direct imaging.
Upper limits from three related studies are plotted in gray (see Section \ref{subsec:limits} for details).\\
Citations: \citet{alcock1996,sumi2011,griest2014,ma2016,barclay2017,mroz2017,hong2018,niikura2019,nielsen2019}.
\label{fig:limit_plot}}
\end{figure*}

If \romanst detects no free-floating planets in a given mass range, it can still place interesting constraints on the occurrence rate of such planets, which in turn can be used to constrain planet formation theories.
We can place expected upper limits on populations of FFPs using Poisson statistics, following \citet{griest1991}.
If we return to our delta function mass distribution such that we assume there is one planet of that mass per star, we can place a 95\% confidence level upper limit for any mass bin, which corresponds to the situation in which we would expect fewer than 3 planets per star \footnote{More specifically, if one expects 3 planets and detects none, according to the Poisson distribution, one could rule out the hypothesis that there are 3 planets at a significance of $1-\exp(-3)\simeq 95\%$. }.
Figure \ref{fig:limit_plot} plots the 95\% confidence level \romanst will be able to place on the {\it total mass} of bodies per star in the MW composed of bodies of mass $M$ if no lenses of that mass are detected. 
Note that the vertical axis is equivalent to $M_p dN/d\log{M_p}$ in units of $M_\oplus$.
For comparison, we plot our fiducial mass function (Equation \ref{eqn:cassan}), and the mass distribution for Solar System bodies\footnote{ssd.jpl.nasa.gov}. 
The latter is to give some intuition as to if there were an equivalent of a Solar System's mass function worth of unbound bodies per star in the MW, but we note that such a mass function is likely to be incomplete at low masses, and possibly also at higher masses~\citep{trujillo2014,batygin2016}. In other words, for typical planetary formation scenarios, a higher number of low-mass objects are ejected than remain in our solar system, and in least a subset of planetary systems, a higher number of higher-mass objects are ejected than remain in our solar system.  

This origin of the shapee of the total mass limit curve deserves some discussion. 
For FFP masses $M\gtrsim1~M_\oplus$ the curve rises as $M^{1/2}$, which is somewhat counter intuitive, though may be recognized by those familiar with dark matter microlensing surveys. 
The number of expected microlensing events \romanst will detect is set by the microlensing event rate $\Gamma$, which scales as the square root of the object mass $\Gamma \propto M^{1/2}$, {\it if there is a fixed number of objects}. 
But the vertical axis of Figure \ref{fig:limit_plot} is the total mass of expected objects of mass $M_p$ per star $M_{\rm tot}$, not the total number. 
So for fixed $M_{\rm tot}$, the number of objects scales as the inverse of the object mass $M^{-1}$ and thus the microlensing event rate produced by a fixed total mass of object scales with the individual object mass as $M^{-1/2}$. 
The total number of detections therefore scales as $N_{\rm det}\propto M_{\rm tot}M_p^{-1/2}$
The survey limit is a contour of a constant number of expected detections, and thus the total mass of ejected objected scales as as $M_{\rm tot} \propto M^{1/2}$. 

Below $M{\sim}1~M_\oplus$, the finite size of a typical \romanst source star becomes larger than the typical Einstein ring radius of the lens, and so the event rate per object becomes independent of object mass. 
But the event rate per total object mass scales as $M^{-1}$, and we would expect the limit curve to become more steeply positive and scale as $M^{-1}$. 
However, the transition to the finite-source dominated regime begins to reduce the peak magnification of events, even if lengthening them, which eventually significantly reduces the probability of a microlensing event being detected. 
Between ${\sim}0.01-1~M_\oplus$, finite-source effects from events with $1<u_0<\rho$ increase the detectable event rate (and reduce the total mass limit) by up to a factor of two relative to events with only $u_0<1$. 
Below $M\lesssim 0.01~M_\oplus$ finite-source effects decrease the maximum magnification of microlensing events to the point where they start to become undetectable, and the detection efficiency begins to fall far faster than the event rate increases, and the slope of the limit curve inverts and becomes sharply negative.

Viewed broadly, the total mass limit curve shows that \romanst will be an extremely sensitive probe of the {\it total mass budget} of loosely bound and free-floating masses. 
At its most sensitive mass, $M\sim3\times 10^{-2}~M_\oplus$ (near the mass of Mercury), \romanst would be sensitive to total masses of just ${\sim}0.1M_\oplus$ per star (or roughly three objects per star).
\romanst will be sensitive to a total mass of $1 M_\oplus$ or less of objects with masses over a range of ${\sim}0.003-100~M_\oplus$, or more than 5 orders of magnitude in mass. While for the lowest mass objects these total masses are large compared to the mass budget of the present Solar System, they are small compared to the total mass of planetesimal disks that are required to form solar-system-like planet configurations in simulations.
For one example, the Nice model considers initial planetesimal disk masses between 30 and 50  $M_\oplus$ beyond Neptune \citep{tsiganis2005}. 
For a broader view of the expected population of loosely bound and free-floating objects, we can compare the \romanst total mass limit curve to various predictions and constraints on these populations.

The first set of comparisons we draw is between \romanst's limits and limits set by microlensing searches for massive halo compact objects (MACHOs). 
There are three studies we consider
\begin{enumerate}
    \item As mentioned in the introduction, \citet{alcock1996} presented combined results of the MACHO and EROS microlensing surveys. 
    These surveys were searching for MACHOs as candidates for the dark matter mass components of the MW halo.
    \item \citet{griest2014} found a similar limit on primordial black holes, but used the \textit{Kepler} transit survey.
    \textit{Kepler} provides relatively high-cadence observations of a fixed, relatively dense, stellar field, which is nearly optimal for a survey of microlensing events. 
    The drawbacks were that this was towards a relatively low stellar density field compared to the LMC or the Galactic Center and that potential sources were much closer than those of typical of microlensing events. 
    The limits placed here are from the analysis of 2 years of the \textit{Kepler} mission, looking for short timescale events. 
    \item \citet{niikura2019} used the Hyper Suprime-Cam on the Subaru Telescope (Subaru/HSC) to perform 2 min cadence observations of M31 with high resolution. 
    This search yielded the best constraint on low mass primordial black holes as a component of the Milky Way Dark Matter halo. 
    \item \citet{niikura2019b} placed limits roughly 50\% lower that than MACHO+EROS result at $50M_\oplus$ using 5 years of OGLE IV data. We do not include this result in Figure 8 due to space constraints.
\end{enumerate}
These limits are not meant to be a direct comparison, so we simply scale their limits by assuming a stellar number density of $n_\star=0.14 \text{~pc}^{-3}$ and a dark matter halo mass density of $\tsub{\rho}{halo}=0.3\text{~GeV}/\text{cm}^{-3}$.
We determine their measured halo mass fractions, $\tsub{f}{HM}$, from their figures. 
Then, the mass of free-floating objects per star is simply 
\begin{equation}
   \frac{\tsub{f}{HM}\tsub{\rho}{halo}}{n_{\star}} = 10^3 \textrm{M}_\oplus\left(\frac{0.06}{\tsub{f}{HM}}\right).
\end{equation}

We also include three frequencies from other observational efforts:
\begin{enumerate}
\item\citet{sumi2011} reported that there may be two free floating Jupiter mass planets per star in the MW. Although inconsistent with the (more recent) limits set by \citet{mroz2017}, we display this result for context.
\item \citet{mroz2017} place an upper limit of fewer than 0.25 Jupiter mass FFPs per star in the MW. 
This is indicated by the black arrow in the upper right. 
\citet{mroz2017} find a tentative signal for five-to-ten 5 $M_\oplus$ mass FFP per star in the MW. 
This is represented by the black vertical bracket.
Note that if no events occur with a Jupiter mass lens, then \romanst~will place a limit of fewer than one Jupiter mass planet per $\sim$100 stars in the MW. 
This will improve the limit placed by the OGLE survey from 8 years of data by more than an order of magnitude \citep{mroz2017}.

\item We consider measurements of the frequencies of \textit{bound} planets and brown dwarfs found using direct imaging by \citet{nielsen2019}.
While these are bound planet frequencies, they are for companions with semi-major axes from 10-100 au, which would likely be mistaken for free-floating planets in microlensing surveys, and thus provide a useful comparison.
\citet{nielsen2019} found a 3.5\% occurrence rate for 5-13 Jupiter mass planets and a much lower rate of 0.8\% for 13-80 Jupiter mass brown dwarfs for hosts with mass $0.2>M/M_\odot>5$.
We include these two frequencies as black circles in Figure \ref{fig:limit_plot} with vertical errors being their reported uncertainties and horizontal the associated ranges. 
\romanst will be sensitive to these widely bound companions, so distinguishing these free-floating planet false positives will be important.
\end{enumerate}

We also plot predictions on the total mass of FFPs per star from a number of theoretical simulations:
\begin{enumerate}
    \item \citet{pfyffer2015} present simulations of formation and evolution of planetary systems, in which only $\sim$0.04 $\tsub{M}{Jup}$ of planets are ejected per star in the optimistic case of no eccentricity or inclination damping.
    \item \citet{ma2016} predict the number of planets ejected per star from dynamical simulations. We take values from their models of $0.3 M_\odot$ stars, in that 12.5\% of stars eject $5M_\oplus$ of mass in $0.3M_\oplus$ bodies.
    \item \citet{barclay2017} predict the number of planetesimals ejected from systems during planet formation. We only compare our limit to their prediction in which giant planets are present in the system, as gray horizontal bar spanning the width of the bins used. In the case that no giant planets are present, fewer objects are ejected. 
    \item \citet{hong2018} predict that $\mathcal{O}(0.01-1)$ moons will be ejected from systems following planet-planet dynamical interactions. We assume these moons have masses from 0.1-1$M_\oplus$, and thus a range of possibilities is included within the gray shaded region. This is a generous upper mass limit compared to the moons of our Solar System, but we note that that little is understood on the formation of exomoons. As an example of an unexpected possibility, there is (contested) evidence of a Neptune-sized exomoon in the Kepler-1625b system \citep{teachey2018,kreidberg2019,teachey2020}.
\end{enumerate}

Thus, we conclude that \romanst~will not only improve the constraints on the abundance of objects with masses from that of less than the moon to the mass of Jupiter by an order of magnitude or more, but it will also allow for a test of model predictions for the total mass of ejected planets in several different planet formation and evolution theories. 

\section{Discussion} \label{sec:disc}

\subsection{Event Detection}
The \romanst~microlensing survey will record nearly $40,000$ photometric data points for $\sim10^8$ stars over its 5 yr duration.
While we have perfect knowledge within these simulations, practically finding events due to very low-mass lenses will likely require more sophisticated search algorithms.
Microlensing surveys have used clear and specific cuts in identifying events.
For example, \citet{mroz2017} made a series of detection cuts based on the temporal distribution of data points during a candidate event, e.g, the number of observations obtained while the flux is rising and falling.  These additional cuts were made in order to avoid false positives like flares or cataclysmic variables.

Machine learning classifiers are also starting to be applied to microlensing survey data as well.
\citet{wyrzykowski2015} searched through OGLE-III data using a random forest classifier.
\citet{godines2019} present a classifier for finding events for low-cadence wide field surveys.
\citet{khakpash2019} developed a fast, approximate algorithm for characterizing binary lens events.
Bryden et al. (in prep.) are developing a machine learning classifier for the microlensing survey being performed with the United Kingdom Infrared Telescope \citep[UKIRT,][]{shvartzvald2017}.
This survey is designed to be a pathfinder for the \romanst~survey, and is mapping the near-infrared microlensing event rate in candidate \romanst fields. 

Still, most of these efforts have focused on the familiar regime of small finite source sizes $\rho\sim 10^{-2}-10^{-3}$ regimes that are most familiar in microlensing surveys.
It will need to be carefully examined how effective these search techniques are in detecting the extremely short-timescale events we are considering, particularly those with qualitatively different morphologies from the more familiar $\rho\ll 1$ single lens microlensing events.
Here we use only the $\Delta\chi^2$ and $n_{3\sigma}$ metrics to determine if an event is detected in these simulations (but see Appendix \ref{sec:det_thresh}). 
However, events may be detectable by \romanst~over a wider region of parameter space using different event selection filters, including the low-amplitude top hat events caused by low mass lenses.

\subsection{False Positives}
\label{subsec:false_pos}
The full sample of microlensing events detected by \romanst will need to be vetted for false positives. 
Detector artifacts such as hot pixels or other defects may introduce systematics that could mimic a short-timescale microlensing event.
However, among the Core Community Surveys, the Galactic Exoplanet Survey provides one of the best opportunities to characterize the Wide Field Instrument H4RG detectors, facilitating the exclusion of these artifacts in light curves \citep{gaudi2019}.

Astrophysical sources could also be mistaken for microlensing events.
Similar to those false positives of ground based microlensing surveys, these will include at least asteroids, cataclysmic variables, and flaring stars.
While a more rigorous event detection algorithm will allow \romanst to mitigate these in the full microlensing sample, even simple cuts will sometimes suffice.
For example, M dwarf flares rarely last longer than 90 minutes \citep{hawley2014} making their exclusion almost guaranteed by a simple cut on $n_{3\sigma}$. 
Longer lasting events such as novae or cataclysmic variables will have many photometric data points during their eruption to model and reject them. 
Spatio-temporal clustering of candidate events and astrometric centroid analysis can be used to recognize asteroids that have not already been identified prior to a pipeline run.

\subsection{Degeneracies}

As identified in \citet{mroz2017}, one event (MOA-ip-01) in the sample of short timescale events presented by \citet{sumi2011} has a degenerate solution.
The event was reported with \tE~$=0.73$ d, but an alternate solution with much longer $t_{\rm E}=8.2$ d is favored. 
In this case, a larger blending parameter ($f_s$) and smaller impact parameter ($u_0$) resulted in the alternate solution. 
The major difference between these models is in the appearance of the wings of the magnification event. 
This degeneracy is well described in \citet{wozniak1997}.
\romanst should be able to distinguish between these approximately degenerate events via its high precision and cadence.

Another relevant degeneracy occurs in lensing events with large relative angular source size $\rho$.  
In this regime, the magnification over the duration of the event is roughly constant (in the absence of limb darkening) and set by Equation \ref{eqn:fse_peak}. 
As a result, $\rho$ becomes nearly degenerate with the blending parameter $\tsub{f}{S}=\tsub{F}{S}/(\tsub{F}{S}+\tsub{F}{B})$, which is the fraction the source flux $\tsub{F}{S}$ contributes to the observed baseline flux $\tsub{F}{S}+\tsub{F}{B}$, where $\tsub{F}{B}$ is the blended flux.
Essentially, the flux from the source alone cannot be confidently measured without precise and dense photometry, which can be used to distinguish the subtle differences in these broadly degenerate light curves.  
Thus, in the presence of blended light, one may underestimate the true peak magnification \citep[][Johnson et al., in prep.]{mroz2020} making it difficult to constrain $\rho$ precisely.  
This is important because $\rho$ depends on the angular source size which will be poorly constrained when no color measurement is made while the source is magnified.
These measurements will likely not occur for short timescale events and those that exhibit extreme finite-source effects. 
We note that this degeneracy persists even in the presence of limb darkening (Johnson et al., in prep.)

Since, for fixed $\theta_*$, $\rho$ increases as the planet mass (and thus the $\tsub{\theta}{E}$) decreases.  Thus as the lens mass decrease, more and more events enter into the $\rho\gg 1$ regime, thereby increasing the likelihood that they will suffer from this degeneracy.
We note that this continuous degeneracy is different than the discrete degeneracy exhibited in the event reported by \citet{chung2017}.

\begin{figure}[t]
\epsscale{\epsScaleFactor}
\plotone{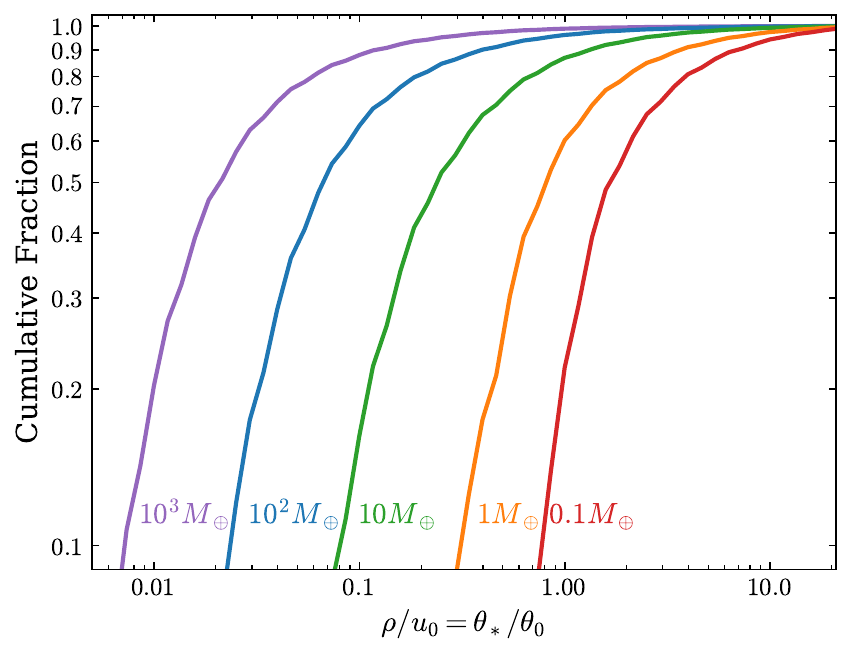}
\caption{
The cumulative fraction of detected events as a function of $\rho/u_0$ which is equal to $\theta_*/\theta_0$, or the angular source radius relative to the angular impact parameter.
Almost all low-mass lens events detected will exhibit finite-source effects \citep[with $\rho/u_0\gtrsim0.5$, ][]{gould1997}.
\label{fig:rho}}
\end{figure}

In order to estimate the fraction of events for which finite source effects should be detectable, in Figure \ref{fig:rho} we show the cumulative fraction of detected events as a function of $\rho/u_0$. Events that have $\rho/u_0=\theta_*/\theta_0\gtrsim0.5$ should exhibit finite-source effects \citep{gould1997}.
Events that satisfy this criterion and have $\rho \gg 1$ will be more susceptible to the above $\rho-\tsub{f}{S}$ degeneracy.

\begin{figure}[t]
\epsscale{\epsScaleFactor}
\plotone{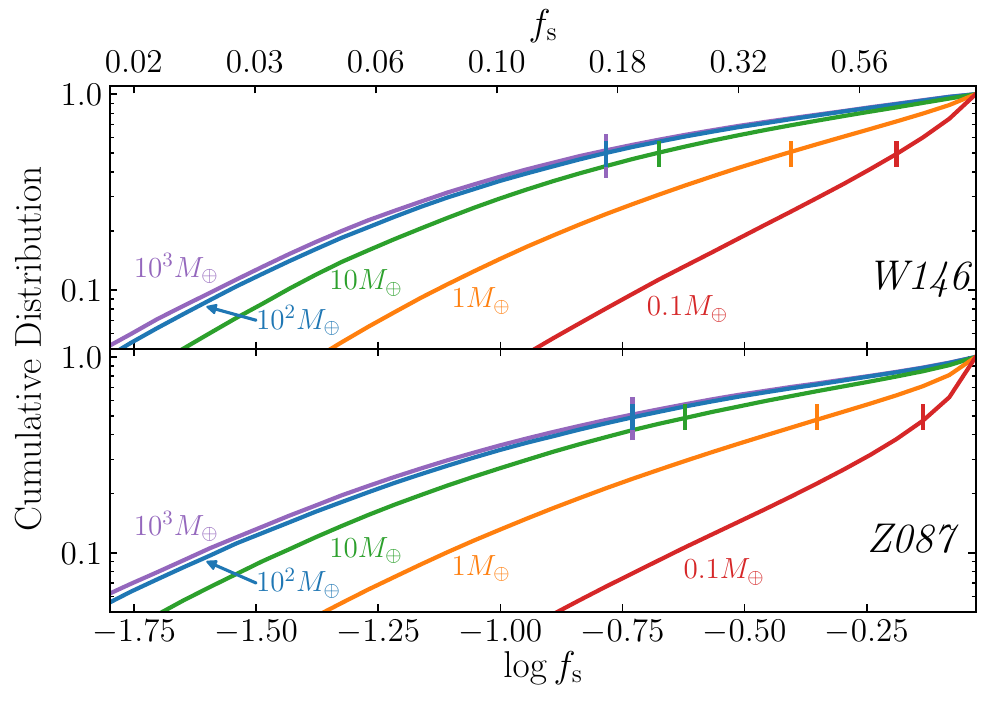}
\caption{
Most source stars will contribute the majority of baseline flux in low-mass lens events.
Normalized cumulative distributions of the blending parameter for detected events among the five mass bins.
For each lens mass, a vertical tick on the distribution marks the value of $f_s$ at which half of events have a greater $f_s$.
For higher mass lenses, this value is $f_s\approx0.20$ and this value only increases as lens mass decreases. 
For the 0.1 $M_\oplus$ lenses, most detected events have $f_s>0.5$ and thus the source makes up the majority of the baseline flux. 
\textit{Upper:} Blending in \textit{W146}.
\textit{Lower:} Blending in \textit{Z087}.
\label{fig:W146_fs}}
\end{figure}

\begin{figure}[t]
\epsscale{\epsScaleFactor}
\plotone{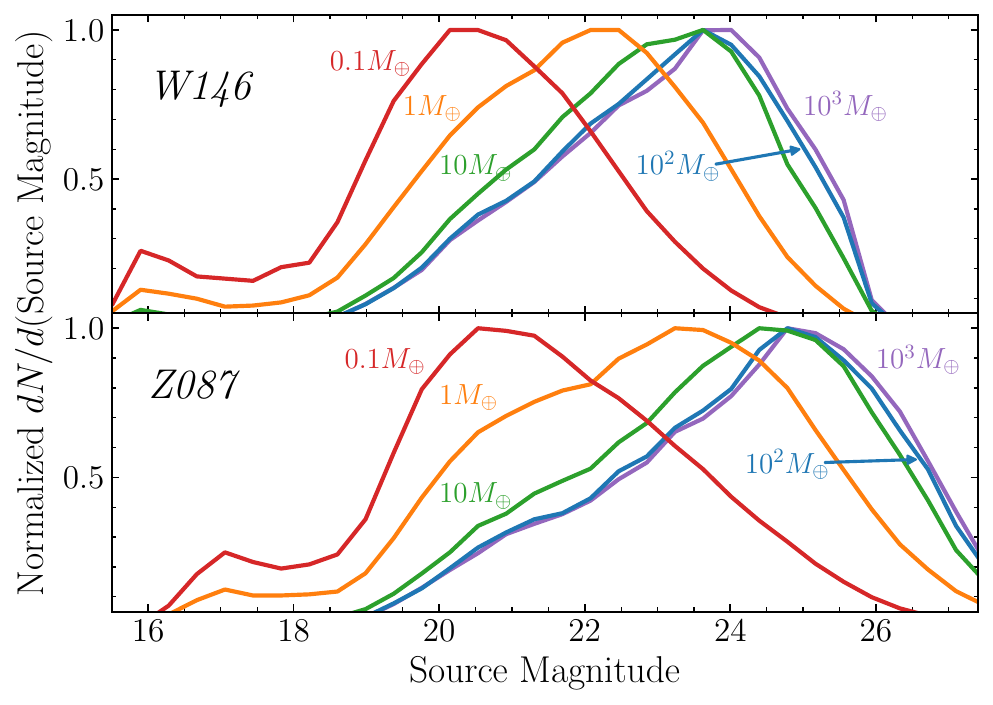}
\caption{
Low-mass lens will have much brighter source stars. 
Here we show the source magnitude distributions normalized to their peaks.
The higher mass lenses ($\gtrsim10M_\oplus$) will have nearly identical distributions, but lower masses than that will strongly deviate. 
\textit{Upper:} Source magnitude distribution in \textit{W146}.
\textit{Lower:} Source magnitude distribution in \textit{Z087}.
\label{fig:source_mag}}
\end{figure}

Fortunately, most of our low-mass lenses that exhibit finite-source effects will be detected in events where the source star dominates the baseline flux (or have large values of $\tsub{f}{S}$). 
This is shown in Figure \ref{fig:W146_fs}, where we plot the cumulative fraction of detected events as a function of $\tsub{f}{S}$ in \textit{W146} (upper panel) and \textit{Z087} (lower panel).
Vertical dashes mark the median values of $\tsub{f}{S}$ of these distributions, and note that the markers for $10^2$ and $10^3M_\oplus$ lie on top of each other.
We also include the source magnitude distributions for detected events in Figure \ref{fig:source_mag} for \textit{W146} (upper panel) and \textit{Z087} (lower panel).
Brighter sources will contribute most to the low-mass lens event rates, as one would expect.  
However, because the fraction of blended flux is not known {\it a priori}, this argument can only be used in statistical sense.

Events of all masses will have little blending in \textit{Z087}, but small mass lenses (that last $\lesssim6$ hr) will be unlikely have a \textit{Z087} measurement taken while magnified.  Measurements from multiple filters may allow an estimate of the source color.  
Table \ref{tbl:color_frac} shows the fraction of events that will have a color measurement taken while the source is magnified, resulting the  breaking of the degeneracy. 
Only 11\% of $0.1 M_\oplus$ lenses will have a color measurement if the \textit{Z087} measurements (or other alternative band) have a cadence of 12 hr, but this fraction can be more than tripled if the cadence increases to 6 hr. 
At 1.0 $M_\oplus$, a larger total number of detected events results in the fraction increase less with an increase in cadence. 
We also include fractions of detected events with color measurements if our threshold $n_{3\sigma}\geq6$ were to be relaxed to only 3, making the percentage even lower for low-mass lenses.
Note the modest decrease in percentages arises from the fact that many more low-mass lens events are detected when the $n_{3\sigma}$ threshold is relaxed (see Appendix \ref{sec:det_thresh}, especially Figure \ref{fig:n_3sig}). 

Alternatively, this degeneracy may be broken through the 5 year baseline of the microlensing survey. 
Potentially blended sources may become apparent as blended stars (either unrelated stars, the host star if the planet is actually bound but widely-separated [see Section \ref{sec:w-b}], or a companion to the host star) move away from the line of sight to the source. 
This fact will also be used in constraining the presence of potential host stars to FFP candidates. 

Johnson et al. (in prep.) demonstrate that there is a second degeneracy in events with $\rho\gg1$.  
This is a multi-parameter degeneracy between the effective timescale of the event, which is well approximated by the time to cross the chord of the source $t_c$, the impact parameter of the lens with respect to the center of the source $u_{0,*}$, and the time to cross the angular source radius $t_*=\theta_*/\mu_{\rm rel}$ (see Section \ref{sec:phenomenology} for definitions of these quantities).  This is easiest to understand in the absence of limb darkening.  A larger impact parameter $u_{0,*}$ results in a shorter event, but with the same peak magnification (due to the `top hat' nature of events with $\rho\gg 1$).  But this shorter duration can be accommodated by scaling $t_*$.  Since neither $\theta_*$ nor $\mu_{\rm rel}$ are known {\it a priori}, it is impossible to measure $\mu_{\rm rel}$ in the regime where these assumptions hold.  Johnson et al. (in prep.) demonstrate that this degeneracy holds for limb darkened sources as well.  

We are investigating the severity and impact of these degeneracies on the ability to recover event parameters in events with extreme finite-source effects (Johnson et al., in prep.).

\begin{deluxetable}{lrrr|rrr}
 \tablecolumns{5} 
\tablecaption{Fraction of detected events with color measurements while source is magnified \label{tbl:color_frac}}
\tablehead{
\colhead{Mass} & \multicolumn{3}{c}{$n_{3\sigma}\geq6$} & \multicolumn{3}{c}{$n_{3\sigma}\geq3$} \\
\colhead{[$M_\oplus$]} & \colhead{12 hr} & \colhead{6 hr }& \colhead{3 hr} & \colhead{12 hr} & \colhead{6 hr } & \colhead{3 hr}}
\startdata
0.1 & $11\% $ & $35\% $ & $56\% $ & $8\% $ & $25\% $ & $42\% $ \\
1.0 & $12\% $ & $23\% $ & $37\% $ & $11\% $ & $20\% $ & $34\% $ \\
10 & $32\% $ & $53\% $ & $75\% $ & $32\% $ & $53\% $ & $75\% $ \\
100 & $74\% $ & $92\% $ & $96\% $ & $75\% $ & $93\% $ & $96\% $ \\
1000 & $98\% $ & $99\% $ & $99\% $ & $98\% $ & $99\% $ & $99\% $ \\

\enddata
\end{deluxetable}

\subsection{Wide-bound Confusion\label{sec:w-b}}

While \romanst~will have sensitivity to the short time scale events of FFPs, true FFPs can be confused with widely separated but bound planets. 
If a bound planet has a large enough projected separation, the source may only be magnified by the planet and not the host star \citep{distefano1999, han2003, han2005}. 
This confusion requires proper accounting if accurate occurrence rates for both FFPs and wide-bound planets are to be reached.
To this end, \citet{han2005} summarize three methods for distinguishing the presence of a host star.

The first method was originally described by \citet{han2003}, in which the magnification by a bound planetary mass object will deviate from that of an isolated lens.
In this scenario, rather than the effective point caustic of an isolated lens, the source is magnified by a planetary caustic which changes the morphology of the light curve.
\citet{han2003} assume some fiducial detection thresholds, and find this method can distinguish $\gtrsim80\%$ of events for projected separations $\lesssim10$au and mass ratios down to $q\approx10^{-4}$.
This deviation was first observed by \citep{bennett2012} and more recently observed in the short (4 day) event reported by \citet{han2020a}, where the presence of a host was determined through 0.03 magnitude residuals near the peak magnification of a single lens model. 

Another pathway to determine if an FFP lens is truly isolated would be to rule out any magnification from a photometrically undetected host \citep{han2005}.
This signal would appear as a long-term, low-amplitude bump in the light curve.
\citet{han2005} show that nearly all planets with projected separations of less than about 13 au will have the presence of their host stars inferred this way.
Assuming the semi-major axis distribution of bound planets is log-uniform, $\sim30$\% of those with $a\in[10^{-1}$,$10^2]$ lie outside 13 au.

The third method is to directly measure blended light from a candidate host.
This can be performed in earlier or later seasons with \romanst by searching for PSF elongation, color-dependent centroid shifts, or event resolution of the lens host star and the source star.
\citet{henderson2016a} find \romanst can exclude hosts down to $\gtrsim 0.1 M_\odot$ depending on the lens distance and the nature of the source star.
\citetalias{penny2019} finds that the majority of hosts to bound planet detections will contribute at least ten percent of the total blend flux. 
The separation of unassociated blended stars (neither the lens or source star) from potential host flux will consider some thought, but could be constrained through priors from the event, such as the distance to the lens system.

\subsection{Limb darkening and wave optics}
\label{subsec:limb}
We did not account for the effects of limb darkening in our simulations \citep{witt1995,herovsky2003}.
The limb darkening profile of source stars is wavelength dependent, with the amplitude of the surface-to-limb variation decreasing $\propto\lambda^{-1}$ for the Sun \citep[][]{hestroffer1998}. 
Because the primary observations are in the near infrared, \romanst\ typical source stars will exhibit less limb darkening than in optical surveys.
As shown by \citet{lee2009} and many more, the limb darkening profile alters the shape of the light curve (see their Figure 6).
This would likely impact our yield estimates for low mass lenses most, where finite-source effects are most likely. 
In our detection cuts, if a source is fainter in its limb, it may shorten the effective timescale of the event. 
This could lower $n_{3\sigma}$ or $\Delta\chi^2$ of the event in our detection threshold. 
However, limb darkening increases the peak magnification of events (see lower panel of Figure \ref{fig:lc_2} which could modestly increase the number of detections we predict. 
We must also consider that \textit{W146} is a wide band and thus the limb darkening will have a significant chromatic dependence over the wavelength range of the filter \citep{han2000, herovsky2003,claret2011}.  

If the mass of a lens is small enough ($\sim10^{-5} M_\oplus$), the geometric optics description of microlensing becomes insufficient and wave effects manifest themselves in the magnification curve \citep[][among others]{takahashi2003}. 
In short, the threshold for this effect is when the wavelength of light being observed becomes comparable to the Schwarzschild radius of the lens; in this limit there is a  fundamental limit to the peak magnification of the event.
For a mass of $10^{-3}M_\oplus$ ($3\times10^{-9}M_\odot$) this corresponds to wavelength of $\sim11\mu m$ (see Equations (5) and (7) of \citet{sugiyama2020}). 
This is below the long wavelength edge of the \textit{W146} band, and the corresponding wavelength only gets longer for larger mass lenses.  We therefore do not consider this effect here.

\subsection{Mass Measurements}
The conversion from timescale to mass for FFP events requires measurements of both the microlensing parallax and the angular Einstein ring. 
A measurement of $\rho$ from finite-source effects and $\theta_*$ from the dereddened source flux and colors\citep{yoo2004} would yield $\tsub{\theta}{E}$, if the degeneracy discussed above can be broken and measurements are made in another filter(s) while the source is magnified\footnote{We note that typically the empirical relations used to convert from source flux and color to angular radius are based on measurements from giant stars, which are most likely to exhibit finite-source effects in microlensing events \citep[e.g.][]{vanbelle1999}.
For lenses with low enough masses, however, we will need appropriately calibrated relations for non-giant source stars \citep[e.g.][]{adams2018}.}.

\textit{Spitzer} enabled the regular measurement of microlensing parallaxes to a large number of stellar, binary, and bound-planetary microlensing events by levering the fact that it was separated by the Earth by $\sim$au due to its Earth-trailing orbit \citep{gould1994,gould1995,gould1999}, but these events had projected Einstein ring sizes of a few au \citep[e.g.,][]{dong2007,yee2015}.
\citet{zhu2016} quantify the potential for simultaneous ground-based observations (and \romanst-only observations) to measure one- and two-dimensional microlens parallaxes. 
Space-based parallax measurements of FFP lenses was also attempted using the Kepler spacecraft during the K2 Campaign 9 survey, which largely consisted of a microlensing survey toward the bulge \citep{henderson2016a,henderson2016b,penny2017,zhu2017a,zhu2017b,zang2018}.
\citet{penny2019b} and \citet{bachelet2019} show that the short intra-L2 baseline between the \textit{Euclid} and \romanst spacecraft would be enough to measure free-floating planet parallaxes.
\citet{ban2020} computes probabilities for measuring parallaxes for combinations of ground and space based telescopes. 
Concurrent observations with wide-field infrared observatories, such as UKIRT \citep{Hodapp2018}, VISTA \citep{Dalton2006}, and PRIME \citep{yee2018}\footnote{\url{https://www.astro.caltech.edu/~srk/\\Workshops/TDAMMS/Files4Facilities/PRIME.pdf}}, as well as wide-field optical observatories, such as DECam \citep{Flaugher2015},  HyperSuprimeCam \citep{Miyazaki2012}, and the Vera C.\ Rubin Observatory \citep{LSST2019}, would enable parallax measurements for both bound and free floating planets.

\section{Conclusion} 
\label{sec:conclusion}

We have used \gulls simulation software \citep{penny2019} to show that {\it Roman Galactic Exoplanet Survey} will inform our understanding of the isolated compact object mass function throughout the Galaxy, down to very low planetary-mass objects.  In particular, it will be able detect microlensing events with timescales as short as 1.5 hr, and thus isolated lenses with masses down to at least 0.1 $M_\oplus$. 
This data set will be used to address questions about both the low mass tail of the initial mass function of stars as well as the total mass and mass function of objects ejected from planetary systems during planet formation and evolution.
\romanst will be able to probe populations of free-floating planets that are essentially impossible to access from ground-based microlensing surveys. 
Finally, the limits that \romanst will place if no such objects are detected would be the most stringent to date by orders of magnitude.

\acknowledgments
We are particularly proud to honor Nancy Grace Roman, after whom this survey telescope has recently been named. 
We hope to live up to her extraordinary influence on space astronomy. 

We appreciate the revisions from the referee that improved the quality of this work, as well as those from careful readings by Radek Poleski and Przemek Mr\'oz. 
We thank our colleagues Andrew Gould and David Bennett for useful discussions. 
We thank everyone on the \romanst Galactic Exoplanet Survey Science Investigation Team. 
We also appreciate Exoplanet Lunch at Ohio State University which was the source of many useful discussions.
SAJ dedicates his contribution to this work to David John Prahl Will, who without this work and that of many others would not be possible.

This work was performed in part under contract with the
California Institute of Technology (Caltech)/Jet Propulsion
Laboratory (JPL) funded by NASA through the Sagan
Fellowship Program executed by the NASA Exoplanet Science
Institute. 
S.A.J, M.T.P., and B.S.G. were supported by NASA grant NNG16PJ32C and the Thomas
Jefferson Chair for Discovery and Space Exploration.

\vspace{5mm}

\software{astropy \citep{astropy2013,astropy2018}, Matplotlib \citep{hunter2007}, MulensModel \citep{poleski2018}, VBBinaryLensing \citep{bozza2010,bozza2018}
          }

\appendix
\section{An introduction to microlensing in the large angular source regime.}
\label{sec:phenomenology}

\subsection{Light curve morphology}

For a typical isolated microlens, the angular size of the source is much smaller than the angular size of the Einstein ring of the lens, and thus the approximation of a point-source generally remains valid.
That is, the magnification as a function of the separation between the source and the lens normalized to the size of the Einstein ring $u$ is given by 
\citep{paczynski1986},
\begin{equation}
    A = \frac{u^2+2}{u(u^2+4)^{1/2}}.
\end{equation}
\label{eqn:apspsl}
The magnification peaks at the minimum separation $u_0=\theta_0/\tsub{\theta}{E}$ where $\theta_0$ is the angular separation between the source and the lens at closest approach. 
The point source approximation in \ref{eqn:apspsl} breaks down when angular separation of the source from the lens, $\theta_0$, becomes comparable to the angular radius of the source star $\theta_*$. 
For point lenses, this condition results in a significant second derivative in the point lens light curve over the angular size of the source, which must be accounted for computing the magnification.  
Thus, for events with impact parameter such that $\rho/u_0 \gtrsim0.5$, where  $\rho=\theta_*/\tsub{\theta}{E}$, the peak of the event (at times $|t-t_0|/\tsub{t}{E} \lesssim 2 \rho$) is affected by finite-source effects \citep{gould1997}.  
Since, for stellar mass lenses, $\rho$ is typically in the range of $10^{-3}-10^{-2}$, most events are unaffected, and those that are effected are high-magnification events.  
Even in such events, finite-source effects are only detectable near the peak of the event, while the magnification during the rest of the event is essentially equivalent to that due to a point source.

However, this characterization breaks completely when the angular size of the source becomes comparable to the angular size of the Einstein ring, or $\rho\gtrsim1$.  
In particular, in the extreme case when $\rho\gg1$, the source will completely envelop the Einstein ring of the lens if it passes within the angular source radius. 
When this happens, the lens magnifies only a fraction of the area of the source as it transits the disk of the source \citep[e.g., ][]{gould1997, agol2003}.
In this limit ($\rho \gg 1$) and to first order, the magnification curve can take on a `top hat' or boxcar shape, saturating at a magnification of
\begin{equation}
    \tsub{A}{peak}\approx1+\frac{2}{\rho^2}
    \label{eqn:fse_peak}
\end{equation}
\citep{liebes1964,gould1997,agol2003}.
The light curve shape is essentially independent of $u_0$ except when $u_0 \sim \rho$ \citep{agol2003}.  Furthermore, the duration of the event is no longer set by the microlensing timescale $\tsub{t}{E}$, but rather is proportional to the source radius crossing time $t_*=\theta_*/\tsub{\mu}{rel}=\tsub{t}{E}\rho$.
\citet{mroz2018a} account for the impact parameter $u_0$ such that they use the time taken for the lens to cross the chord of the source 
\begin{equation}
    t_c = \frac{2\theta_*}{\tsub{\mu}{rel}}\sqrt{1-\left(u_{0,*}\right)^2},
    \label{eqn:t_c}
\end{equation}
where $u_{0,*}=\theta_0/\theta_*$.
Note that this timescale is independent of the angular radius of the Einstein ring and thus lens mass.  
We note that \citet{mroz2017} define $t_*$ as the crossing time over the chord of the source, but we define $t_*$ is the source radius crossing time defined above \citep[see Appendix A of][]{skowron2011}.
Henceforth, we will use $t_c$ for the chord crossing time and we propose that this become the convention.

\begin{figure*}
\epsscale{\epsScaleFactor}
\plotone{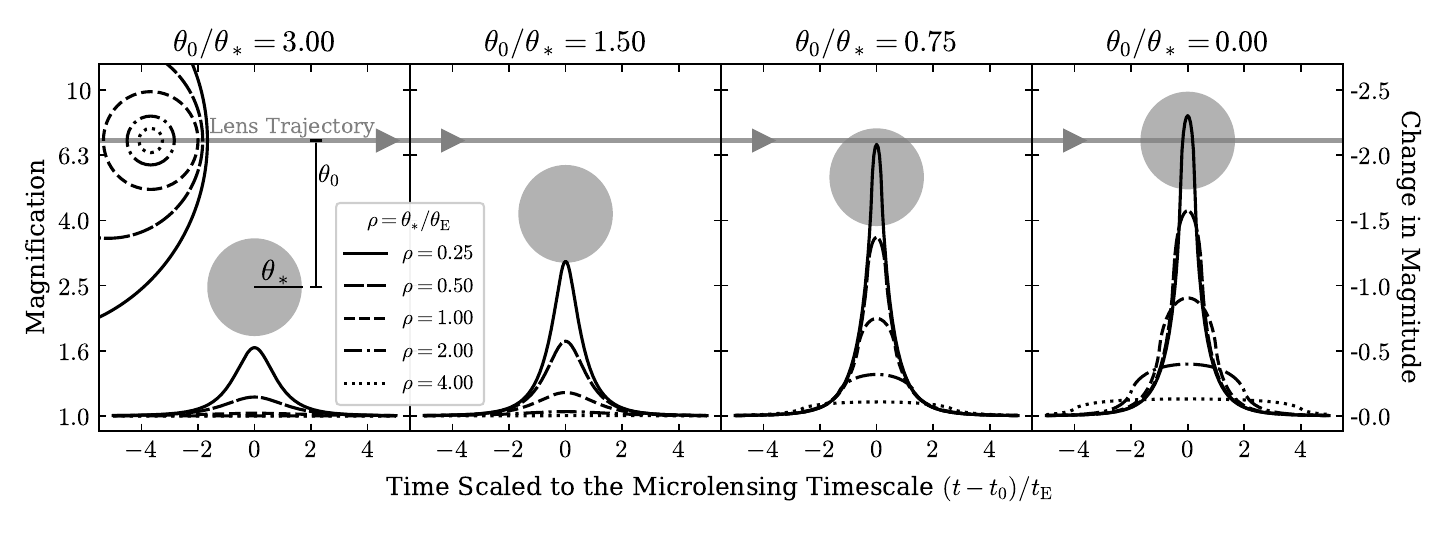}
\caption{The morphology of microlensing light curves changes as finite-source effects become more prominent. 
In the background, we show a gray circle that represents the source (with an angular radius $\theta_*$). 
We also show five Einstein rings scaled to the source size, which have $\rho=\theta_*/\tsub{\theta}{E}$ values indicated in the legend.
Each panel is for a different impact parameter $u_{0,*}=3.00$, 1.50, 1.00, and 0.00 from left to right.
We change the scaled position of the source star circle relative to the lens trajectory (gray horizontal line) to match the impact parameter. 
For each panel, we plot the magnification as a function of time scaled to the microlensing timescale $(t-t_0)/\tsub{t}{E}$.
For the most extreme case of $\rho=4.00$, we see no appreciable magnification until the lens traverses the source ($u_{0,*}<1$) at which point the magnification is essentially constant (except when the lens is near the edges of the source). The light curves thus have a `top hat' appearance. 
We note that this `top hat' morphology only appears when there is no limb darkening.
All events have peak magnifications that monotonically increase $u_*$ decreases, however, this maximum magnification begins to saturate at the expected value of $1+\frac{2}{\rho^2}$ for $\rho>1$.
However, the length of those events with $\rho>1$ are significantly longer than expected from their analytic $\tsub{t}{E}$ timescales. 
\label{fig:fse}}
\end{figure*}

Broadly, these changes in light curve morphology are referred to as extreme finite-source effects, as the light curve is affected by finite-source effects throughout the duration (e.g., at no time while the source is magnified does the point source approximation hold).
We demonstrate the impact of FSEs on the light curve of events in Figure \ref{fig:fse}.
We consider five lenses in which we only vary the angular size of the Einstein ring, quantified by $\rho=0.25$, 0.50, 1.00, 2.00, and 4.00 (as the size of the source is fixed).
The sizes of these rings are shown in the upper left corner of the leftmost panel scaled to the size of a source star, which is depicted as a gray circle (with an angular radius of $\theta_*$).
A horizontal, gray line depicts the path the lenses will take, and it is separated from the center of the source star by the impact parameter $\theta_*$ (again, \textit{to scale}).
We then vary the impact parameter from $u_{0,*}=3.0$ down to $u_{0,*}=0.0$ from left to right.
This is written above the plot and depicted as the source star (gray circle) approaching the lens trajectory (gray line).
Note that the circles and their separations are independent of the time and magnification axes.

Each panel depicts all five lenses in different events, where the line-style matches the lens with the same Einstein ring line-style in the upper left corner of the leftmost panel.
We use the method of \citet{lee2009} as implemented in \texttt{MulensModel} \citep{poleski2018} to compute all light curves in this figure. 

As $\tsub{\theta}{E}\propto\sqrt{M}$, the more massive lens will have the largest $\tsub{\theta}{E}$ and be the furthest from the regime of finite-source effects. 
Note that time is referenced to the peak of the event ($t_0$), and scaled by the analytic timescale $\tsub{t}{E}$ on the horizontal axis.
The solid-line light curve behaves essentially how you expect an isolated lens to behave as the impact parameter drops up until the last two panels where the peak becomes more rounded. 
This is the first breakdown we described that occurs when $\rho/u_0\gtrsim0.5$, or when the size of the source star is within a few times the impact parameter.

However, the behavior is dramatically different for the lowest mass lens ($\rho\gg 1$).
In this case, when lens is not transiting the source ($u_*>1$), there is effectively no magnification. 
For the smallest impact parameter (rightmost panel), the lightcurve looks like the top-hat described earlier, magnifying the source by roughly 10\% ($1+\frac{2}{4^2}\approx1.13$).
Also note the duration of this event is now much longer than one would expect given its analytic timescale $\tsub{t}{E}$. 
In fact when $u_{0,*}=0.00$, the duration is nearly exactly what we predict given the diameter crossing time $t_c/\tsub{t}{E}=2\rho=4$ for the $\rho=2$ case and $t_c/\tsub{t}{E}=8$ for the $\rho=4$ case.  
In the rightmost panel, the $\rho=4$ event lasts $\sim4$ times longer than one would expect based on the value of $\tsub{t}{E}$.

To provide a quantitative sense of the relevant scales, consider a typical stellar mass lens ($0.3 M_\odot$), which has an angular Einstein ring radius of $\tsub{\theta}{E}=550~\mu$as.
A source star in the Galactic bulge (at a distance of $\tsub{D}{S}=8$ kpc) that has a radius of 1 $R_\odot$ will have an angular radius of just $0.6~\mu$as.
Lenses with mass $\lesssim0.12M_\oplus$ will have $\rho\gtrsim1$ for this source. 
A typical clump giant in the bulge will have a radius of $\sim10R_\odot$, leading to $\rho>1$ for lenses with mass $\lesssim10M_\oplus$.

These morphological changes will impact the microlensing event rate and microlensing optical depth \citep{vietri1983,paczynski1991}.
Recall that the microlensing optical depth (the probability any given star is being lensed) is a function of the fraction of the sky covered by Einstein rings. 
As demonstrated above, lenses with small enough masses will have Einstein rings smaller than the angular size of some stars. 
\citet{han2005} show that for lenses with low enough masses, the event rate actually \textit{increases} compared to what you would expect for lenses with Einstein rings smaller than the angular size of source stars.
For these lenses, the event rate is proportional to the fraction of the sky covered by source stars.
However, the detection of such events is hampered by the fact that the peak magnification is lower than one would expect for a point source. 
\citet{han2005} also derive analytic expressions for the threshold impact parameter for detection and the minimum detectable mass lens as function of the threshold signal-to-noise ratio for detection. 

In reality, the shape of the light curve is sensitive to the limb-darkening profile of the source as well as any of its surface features \citep[e.g.][]{witt1994,gould1996,agol2003,herovsky2003,yoo2004,lee2009}.
The impact of included limb darkening is shown in the lower panel of Figure \ref{fig:lc_2}.
The `top hat' shape disappears, and the light curve becomes more rounded. 
The example in Figure \ref{fig:lc_2} adopted a single parameter linear limb darkening profile, but more structure could be added if a more complex profile was used \citep[e.g.,][]{claret2011}, or if surface features (such as star spots) were considered \citep{herovsky2003}. 

\section{Detection Criteria}
\label{sec:det_thresh}
We require that simulated events pass two criteria to qualify as detections.
The first is based on the deviation ($\Delta\chi^2$) the event causes from a flat light curve
\begin{equation}
\Delta\chi^2= \tsub{\chi}{Line}^2 -\tsub{\chi}{FSPL}^2
\end{equation}
where $\tsub{\chi}{Line}^2$ is the $\chi^2$ value of the simulated light curve for a flat line at the baseline flux and $\tsub{\chi}{FSPL}^2$ is the same but for the simulated data to the injected finite-source point-lens model of the event.  The second criterion is the number of consecutive data points that are measured $3\sigma$.  In this section we isolate the effect of the value of each criterion on the yield as a function of planet mass in turn, and then consider the complex interplay between them.  

\begin{figure}[t]
\epsscale{\epsScaleFactor}
\plotone{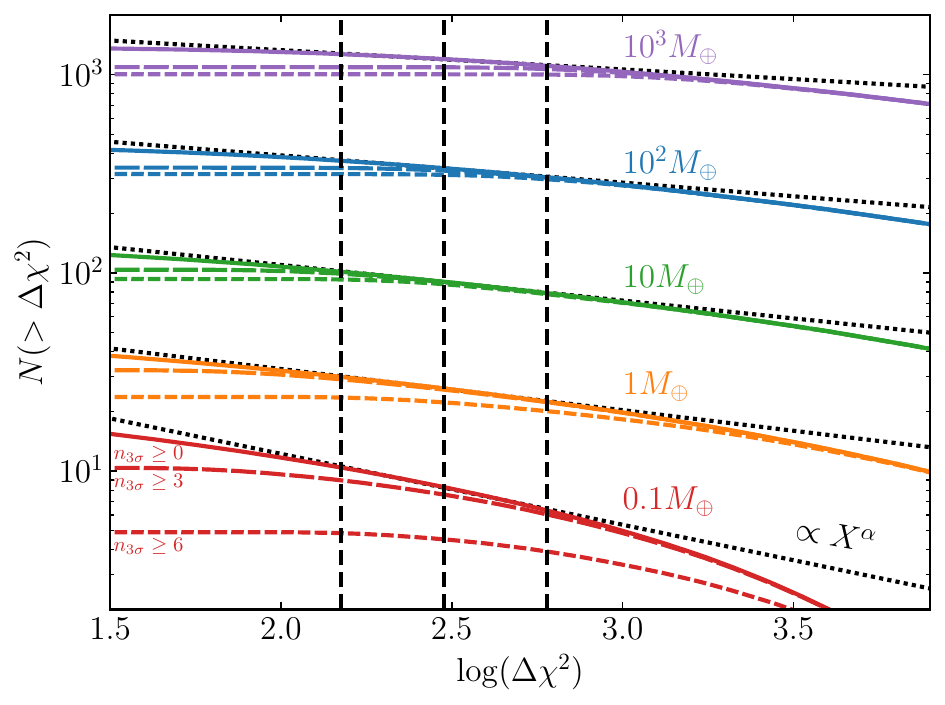}
\caption{The cumulative distribution of the $\Delta\chi^2$ of simulated events.
From bottom (red) to top (purple), the solid lines represent yields of lenses with masses of $0.1M_\oplus$ to $10^3M_\oplus$, assuming one FFP of that mass per star in the MW.
From left to right, the vertical, black dashed lines indicate where $\Delta\chi^2=150$, $\Delta\chi^2=300$, and $\Delta\chi^2=600$.
We fit a power law (Eqn. \ref{eqn:chi_power}) to each of the solid lines for a range of $\Delta\chi^2\in[150,600]$. 
The slopes of these are included in Table \ref{tbl:chi2_slopes}.
We also plot the cumulative $\Delta\chi^2$ distributions for events with $n_{3\sigma}\geq3$ and $\geq6$ as the long-dashed and dashed lines, respectively.
The distributions flatten significantly for lower $\Delta\chi^2$ when we require $n_{3\sigma}\geq6$.
\label{fig:n_chi2}}
\end{figure}

We first plot the cumulative number of detected events $\tsub{N}{det}(X\geq\Delta \chi^2)$ as a function of the threshold $\Delta \chi^2$ in Figure \ref{fig:n_chi2}.
We show the cases of our five discrete masses under the assumption that there are one such planet per star. 
Applying only this criterion, we can analytically estimate the impact of mission/survey design changes on the yield of FFPs by inferring the impact those changes would have on the $\Delta\chi^2$ of events \citepalias[akin to ][]{penny2019}.
This is because the distributions in Figure \ref{fig:n_chi2} can be locally fit by a power law
\begin{equation}
N(\Delta\chi^2>X)\propto X^{\alpha},
\label{eqn:chi_power}
\end{equation}
over a wide range of $\Delta\chi^2$, as has previously been shown by \citet{bennett2002}. 
We fit this power law for each mass on the range $\Delta\chi^2=[150,600]$, and list the values of exponent in Table \ref{tbl:chi2_slopes}.

While we necessarily employ a $\Delta\chi^2\geq300$ as one of our thresholds, basing detection rates solely on this criterion is problematic because of the potential for very short events, e.g., events
with only a few extremely magnified points that together contribute more than 300 to the $\Delta \chi^2$.  
Modeling these events would be challenging.  
We therefore also impose the second criterion on $n_{3\sigma}$, which is specifically the number of consecutive data points with $n>3\sigma$ above the baseline flux of the source star plus blend\footnote{We note that a similar criterion were used by \citet{sumi2011} and \citet{mroz2017}}. 
This criterion ensures that there will be a sufficient number of high signal-to-noise ratio data points during the events that they can be confidently modelled.

To illustrate how this criterion changes the cumulative number of detections relative to just applying the $\Delta\chi^2$ criterion, in Figure \ref{fig:n_chi2} we plot as dashed (long-dashed) lines the distributions also requiring $n_{3\sigma}\geq6$ ($n_{3\sigma}\geq3$) points.
We fit the slopes of the cumulative distributions as power laws distributions as before over the same range, and include the power law exponents in Table \ref{tbl:chi2_slopes}.
For a given mass, the distributions we derive applying both criteria change relative to only applying the $\Delta\chi^2$ criterion in a manner that depend on the mass of the lens. 

We note that, for all of the masses, the cumulative distributions begin to fall below the power law fit to the solid curves (without the $n_{3\sigma}$ cut) at the highest values of the threshold $\Delta\chi^2$.  
Furthermore, the onset of this deviation occurs for lower values of $\Delta\chi^2$ for the very smallest masses. 
This deviation is due to the onset of finite-source effects, and the increasing importance of these effects for lower masses.  

Conversely, for lower values of the threshold $\Delta\chi^2$, the cumulative distributions begin to fall below the power law fit to the solid curves at roughly the same value of $\Delta\chi^2$ for the three largest masses, but at different values for the lowest two masses.
Finally, for all the masses, the cumulative distribution of the number of detections becomes essentially flat (independent of the $\Delta\chi^2$ threshold) for values of $\Delta\chi^2\la 150$ and $n_{3\sigma}\geq 6$.  
Thus, for this combination of detection criteria, the yield does not improve with a lower $\Delta\chi^2$ threshold, only with changing the $n_{3\sigma}$ cut.  
These behaviors are all consistent with expectations based on the gradual change in the morphologies of the light curves as finite-source effects begin to dominate (roughly for masses $\la M_\oplus$ (See Section \ref{sec:phenomenology}).

\begin{deluxetable}{lrrr}
 \tablecolumns{3} 
\tablecaption{Slopes of $\Delta\chi^2$ distributions\label{tbl:chi2_slopes}}
\tablehead{
\colhead{($M_\oplus$)} & \multicolumn{3}{c}{$\alpha$} \\
\colhead{} & \colhead{$n_{3\sigma}\geq0$} & \colhead{$n_{3\sigma}\geq3$}& \colhead{$n_{3\sigma}\geq6$}
}
\startdata
0.1 & $-0.36$ & $-0.28$ & $-0.15$ \\
1 & $-0.21$ & $-0.19$ & $-0.12$ \\
10 & $-0.18$ & $-0.17$ & $-0.13$ \\
100 & $-0.14$ & $-0.080$ & $-0.040$ \\
1000 & $-0.097$ & $-0.013$ & $-0.0020$ \\

\enddata
\end{deluxetable}  

\begin{figure}[t]
\epsscale{\epsScaleFactor}
\plotone{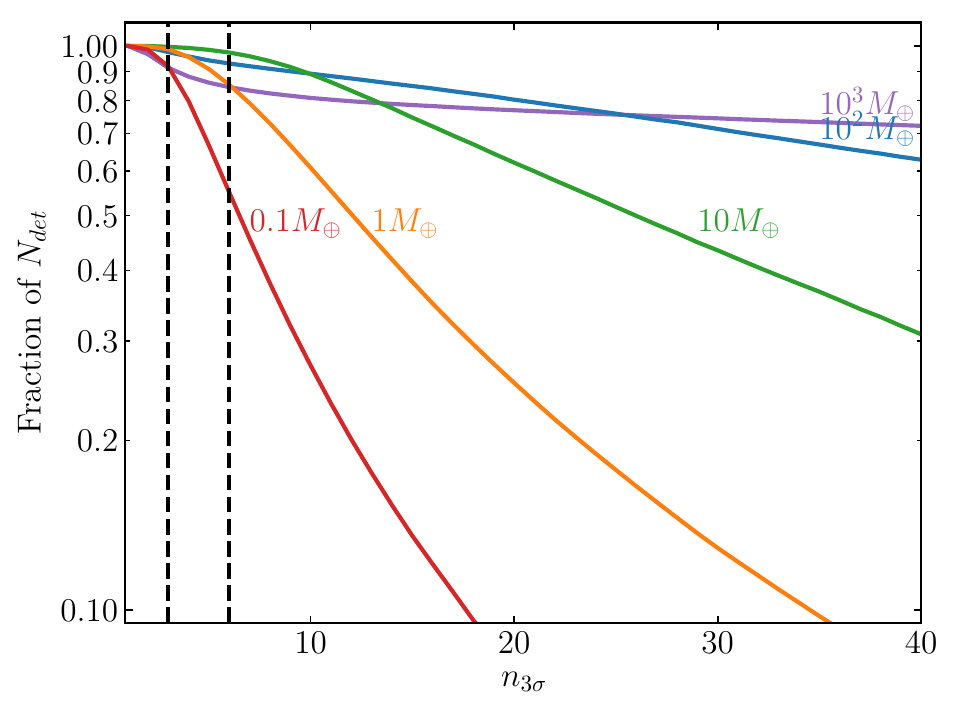}
\caption{The cumulative fraction of events as a function of $n_{3\sigma}$ with $\Delta\chi^2>300$.
Each line represents only events with the labeled mass. 
The left (right) vertical dashed lines are at $n_{3\sigma}=3$ $ (n_{3\sigma}=6)$.
The most significant difference between these thresholds is for the yields of $0.1M_\oplus$ FFPs, which nearly doubles when the threshold is relaxed.
These events are typically short or have low magnification. 
For masses above $10M_\oplus$, the number of detections is relatively independent over the range $6<n_{3\sigma}<40$ . 
We conclude that the impact of our $n_{3\sigma}$ detection criterion is highest for low-mass lensing events.\\
\label{fig:n_3sig}}
\end{figure}

To further explore the interplay between the two detection criteria, we isolate the effect of the $n_{3\sigma}$ cut on the yields in Figure \ref{fig:n_3sig}. Here we show the cumulative fraction of events as a function of $n_{3\sigma}$ for events with $\Delta\chi^2\geq300$. 

For the two largest masses, the yield is a relatively weak function of $n_{3\sigma}$ since these masses typically give rise to longer timescale (and thus more well sampled) events.  
Interestingly, we find that 10$M_\oplus$ events are the most robust to this selection criterion for $n_{3\sigma}\lesssim10$, however it falls off quickly afterwards, as expected.  
The lowest two masses continue this trend, become ever more sensitive to the value of the $n_{3\sigma}$ cut at a fixed threshold of $\Delta\chi^2\geq300$.
Again, this is expected as the timescale distributions for the lower and lower masses are typically shorter and shorter compared to the cadence of 15 minutes.

Thus we find that there an important and complex interplay between both these criteria, which makes predicting the impact of changes in yield at different values of the photometric precision at a given magnitude more difficult than if we just imposed the  $\Delta\chi^2$ threshold.
As a concrete example to illustrate this point, imagine an event with 5 data points 8$\sigma$ above the baseline flux and the next most significant point being only 3$\sigma$ above baseline.  Further assume that all these points are consecutive, and together yield a total $\Delta\chi^2=329$.
We could change our threshold to $\Delta\chi^2\geq329$, and our event would still be counted as detected (at it still passes the $n_{3\sigma}\geq6$ cut). 

As discussed previously, changing the threshold in $\Delta\chi^2$ is equivalent to scaling the photometric precision of the survey as function of magnitude. However, simply scaling the yield with the threshold $\Delta\chi^2$ doesn't capture the impact on the $n_{3\sigma}\geq6$.  
What is really of interest is how the number of detected events when we rescale the individual uncertainties including both criteria.  This demonstrates how robust the yield is to degradation or improvement in the photometry.  In the above example, assume the uncertainties are increased by $\sim4.7\%$, such that the event now has $\Delta\chi^2=300$ and thus would still (barely) pass the $\Delta\chi^2$ criterion for detection. 
However, the last point would now have a significance of $\sim2.9\sigma$, and thus the event would fail our $n_{3\sigma}$ criterion, and no longer be detected.
This means that those distributions in Figure \ref{fig:n_chi2} and \ref{fig:n_3sig} can only be used to  predict the change in the expected number of detections resulting from changes in the two detection criteria, but cannot be used to assess the impact on the yield if the photometric precision changes at fixed magnitude.  The latter is of more interest when estimating the changes in the survey yield as a result of changes in the mission design. 

\begin{figure}[ht]
\epsscale{\epsScaleFactor}
\plotone{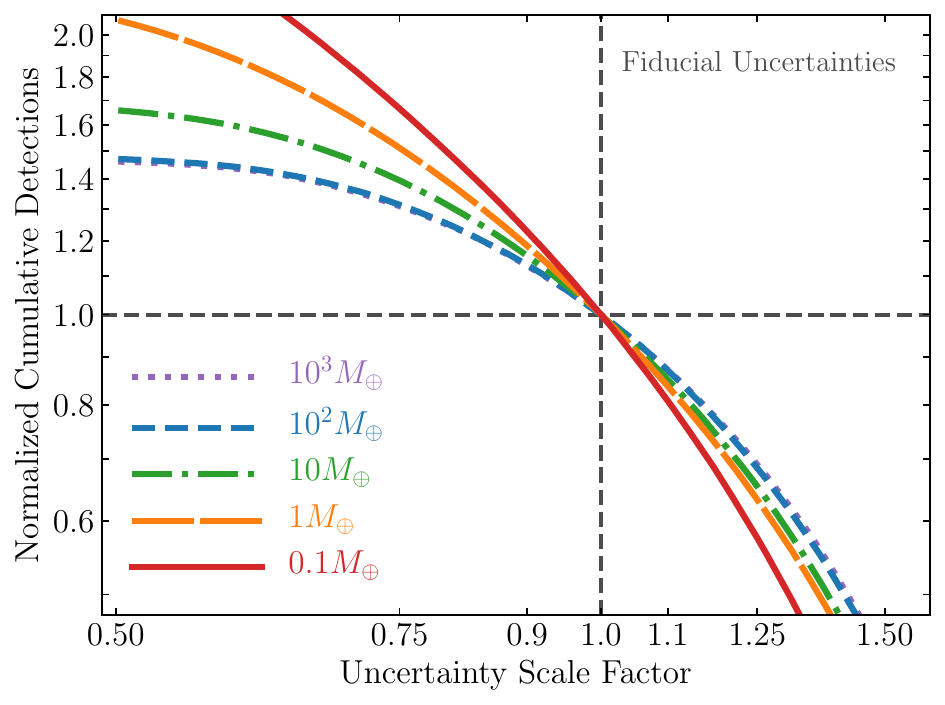}
\caption{
The number of low-mass lens detections that pass both our detection criteria has a higher dependence on the \romanst photometric precision than high-mass lenses.
These distributions are normalized to the number of events in each mass bin detected when no scaling is applied.
The distributions for events with $10^2$ and $10^3$ $M_\oplus$ behave as though only the $\Delta\chi^2$ threshold were being scaled, as these events have long timescales and thus are generally robust to the $n_{3\sigma}$ threshold.
However, as lens mass decreases, the slopes of the curves for the lower masses events continues to steepen for higher error scalings. 
These masses naturally produce shorter timescale events, which are much more susceptible to cuts in $n_{3\sigma}$.
Nevertheless, when considering our detection criteria, these distributions show that \romanst's yield of low-mass lensing events will degrade gracefully with increasing (fractional) photometric precision.
\label{fig:del_n_3sig}}
\end{figure}

Thus, to further investigate this interplay, we ran a separate set of modified simulations where everything is the same as described in Section \ref{sec:sims}, except that we added two calculations. 
For every event, we calculated the factor by which the uncertainties would need to be uniformly scaled by in order that the total $\Delta\chi^2$ of the event is equal to 300, specifically $\tsub{C}{DC2} = ({\Delta\tsub{\chi}{True}^2}/300)^{1/2}.$
We then find the data point that would be the last to qualify the event for $n_{3\sigma}\geq6$ cut, and calculate the factor that the photometric uncertainty of that data point would need to be scaled to reach a $3\sigma$ significance, $\tsub{C}{N3S}$.
The lesser of these two factors is the more stringent criteria, and we can therefore assess how the yield changes when the photometric uncertainty is changed including the impact of both both criteria. 
We find the cumulative distribution of detections as a function of the minimum of these two scaling factors and call it the `Uncertainty Scale Factor'$=\min(\tsub{C}{DC2},\tsub{C}{N3S})$.
We plot this distribution normalized to the number of detections expected when the error bars are not scaled at all for our five reference masses in Figure \ref{fig:del_n_3sig}. 

As one would expect, the larger mass lenses ($10^3$ and $10^2 M_\oplus$ have the essentially the same dependence on this scaling factor.  
These events typically last longer, so the $n_{3\sigma}=6$ criterion is generally not approached.
For both these masses, the behavior is thus the same and is similar to just scaling $\Delta\chi^2$.
As the mass of lenses drops, the distributions begin to increasingly steepen from 10 to 1 to 0.1 $M_\oplus$ (the green dot-dash, orange long dash, and red solid lines). 
As events become shorter and shorter with decreasing mass, more events go undetected due to the consideration of $n_{3\sigma}$ when the uncertainties are increased.
Fortunately, the inverse is also true.
In fact, a larger fraction of events are recovered when the uncertainties are smaller than expected (the spread is larger between the distributions for scale factors less than unity).

The most important takeaway from Figure \ref{fig:del_n_3sig} is that the number of detections is fairly robust to the precise \romanst photometric uncertainties that are achieved across a broad range in lens masses, and there are no "thresholds" in the photometric precision below which the detection rate drops precipitously. 
As a concrete example, for the mostly highly impacted lens mass $0.1M_\oplus$, if precision is 10\% greater than expected we still recover $\sim80\%$ of events.

\end{document}